%% file: main.tex
\documentclass[journal=jctc,manuscript=article]{achemso}

\usepackage[version=3]{mhchem} 
\usepackage{hyperref}
\usepackage{amsmath}
\usepackage{amsthm}
\usepackage{amssymb}
\usepackage{mathrsfs}
\usepackage{graphbox}
\usepackage{subcaption}
\usepackage{tabularx}
\usepackage{nameref}

\usepackage{color}
\usepackage{enumitem}
\usepackage[normalem]{ulem}
\usepackage{comment}

\input{macros}

\input{main_sections/author_info}

\title[]
  {Thermodynamic Transferability in Coarse-Grained Force Fields using Graph Neural Networks}

\abbreviations{AA, CG, MD, ML}

\begin{document}

\begin{tocentry}

\includegraphics[]{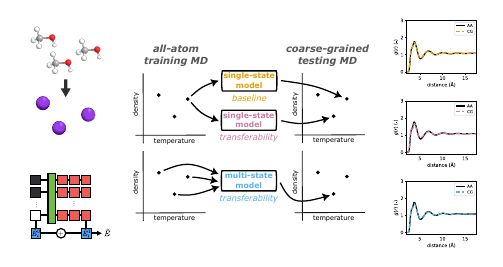}





\end{tocentry}

\begin{abstract}
Coarse-graining is a molecular modeling technique in which an atomistic system is represented in a simplified fashion that retains the most significant system features that contribute to a target output, while removing the degrees of freedom that are less relevant.
This reduction in model complexity allows coarse-grained molecular simulations to 
reach increased spatial and temporal scales
compared to corresponding all-atom models.
A core challenge in coarse-graining is to construct a force field that represents the interactions 
in the new representation 
in a way that preserves the atomistic-level 
properties. Many approaches to building coarse-grained force fields have limited transferability between different thermodynamic conditions as a result of averaging over internal fluctuations at a specific thermodynamic state point.
Here, we use a graph-convolutional neural network architecture, the Hierarchically Interacting Particle Neural Network with Tensor Sensitivity (HIP-NN-TS), to develop a highly automated training pipeline for coarse grained force fields which allows for studying the transferability of coarse-grained models based on the force-matching approach. We show that this approach not only yields highly accurate force fields, but also that these force fields are more transferable through a variety of thermodynamic conditions.
These results illustrate the potential of machine learning techniques such as graph neural networks to improve the construction of transferable coarse-grained force fields.
\end{abstract}


\input{main_sections/introduction}
\input{main_sections/methods}

\input{main_sections/results}
\input{main_sections/conclusion}
\input{main_sections/data}

\begin{acknowledgement}

We acknowledge support from the Los Alamos National Laboratory (LANL) 
Directed Research and Development funds (LDRD).
This research was performed in part at the Center for Nonlinear Studies (CNLS) at LANL. 
This research used resources provided by the Darwin testbed at LANL which is funded by the Computational Systems and Software Environments subprogram of LANL's Advanced Simulation and Computing program. LANL is operated by the Triad National Security, LLC, for the National Nuclear Security Administration of the U.S. Department of Energy (contract no. 89233218NCA000001).

We would like to thank David Rosenberger and Kipton Barros for helpful discussions \ADD{and the referees for their thoughtful reviews.}

\end{acknowledgement}

\begin{suppinfo}

Further details on \ADD{neural network hyperparameter selection,} repulsive potential parameter selection, model training statistics, a discussion on training set size, results of an interpolation test\ADD{, and additional RDFs and ADFs}. \ADD{This information is available free of charge via the Internet at http://pubs.acs.org.}

\end{suppinfo}

\clearpage 

\bibliography{bib/j,bib/CG,bib/misc}

\end{document}


\input{si_sections/hyperparameter}
\input{si_sections/repulsive_potential}
\input{si_sections/training_statistics}
\input{si_sections/data_size}
\input{si_sections/interpolation}
\input{si_sections/addl_rdfs}

%% file: macros.tex
\newcommand{\ourmethod}{{ML CG }} 

\newif\ifHighlitedChanges
\def\ifHighlitedChanges{\iffalse}
\ifHighlitedChanges
  \def\ADD#1{{\color{red}#1}}
  \def\STRIKE#1{{\color{red}\sout{#1}}}
\else
  \def\ADD#1{#1}
  \def\STRIKE#1{\relax}
\fi

\newif\ifHighlitedChangess
\def\ifHighlitedChanges2{\iffalse}
\ifHighlitedChangess
  \def\ADDD#1{{\color{red}#1}}
  \def\STRIKEE#1{{\color{red}\sout{#1}}}
\else
  \def\ADDD#1{#1}
  \def\STRIKEE#1{\relax}
\fi

%% file: main_sections/author_info.tex
\author{Emily Shinkle}
\affiliation[lanlccs]{Computer, Computational, and Statistical Sciences Division, Los Alamos National Laboratory, Los Alamos, NM 87545, USA}
\email{eshinkle@lanl.gov}

\author{Aleksandra Pachalieva}
\affiliation[lanlees]{Earth and Environmental Sciences Division, Los Alamos National Laboratory, Los Alamos, NM 87545, USA}
\alsoaffiliation[lanlcnls]{Center for Nonlinear Studies, Los Alamos National Laboratory, Los Alamos, NM 87545, USA}

\author{Riti Bahl}
\affiliation[emory]{Department of Mathematics, Emory University, Atlanta, GA 30322}
\alsoaffiliation[lanlt]{Theoretical Division, Los Alamos National Laboratory, Los Alamos, NM 87545, USA}

\author{Sakib Matin}
\affiliation[lanlt]{Theoretical Division, Los Alamos National Laboratory, Los Alamos, NM 87545, USA}

\author{Brendan Gifford}
\affiliation[lanlt]{Theoretical Division, Los Alamos National Laboratory, Los Alamos, NM 87545, USA}

\author{Galen T. Craven}
\affiliation[lanlt]{Theoretical Division, Los Alamos National Laboratory, Los Alamos, NM 87545, USA}

\author{Nicholas Lubbers}
\affiliation[lanlccs]{Computer, Computational, and Statistical Sciences Division, Los Alamos National Laboratory, Los Alamos, NM 87545, USA}
\email{nlubbers@lanl.gov}

%% file: main_sections/introduction.tex
\section{Introduction}

Molecular simulations elucidate the microscopic physical processes that give rise to a physical system's function and behavior. 
One of the principal components that determines the accuracy of a molecular simulation 
is the force field, a mathematical model that calculates the forces acting on the particles in the system as a function of their positions,
i.e., the interatomic forces. Force fields are typically constructed and calibrated by a combination of top-down parameterization techniques (so that a simulation reproduces known properties of the target system such as structural, thermodynamic, and dynamical properties measured in experiments) and bottom-up techniques (by fitting to forces generated using first-principles calculations~\cite{Voth2005, Lu2021OPLS, Unke2021, Poleto2022, Zongo2022first, Matin2024}).
Molecular models with well-parameterized force fields enable the determination of key physical and chemical properties needed by researchers in a variety of domains such as chemistry, materials science, and biophysics~\cite{Ramprasad2017, Wu2023, Unke2021, Rosenberger2021}. These properties are generated by extracting observable quantities from the results of a simulation using a variety of sampling techniques~\cite{FrenkelSmit96,Fedik2022}.




Even with a force field in hand, performing molecular simulations can incur a significant computational cost, primarily because large numbers of atoms are often required in order to understand a system's collective behaviors and statistical properties, \ADDD{and timescales associated with processes of interest can be far longer than the timescales on which atoms themselves evolve. Many techniques have been developed to bridge this gap between atomistic and experimental scales, for example, Accelerated Molecular Dynamics\cite{perez2009accelerated}, enhanced sampling methods such as metadynamics\cite{henin2022enhanced}, and correlation function approaches to thermodynamic information such as using the Bogoliubov–Born–Green–Kirkwood–Yvon (BBGKY) hierarchy\cite{Rudzinski2014}.} To ameliorate the computational cost associated with simulating a system at an all-atom (AA) level, the system can be redefined using a coarse-grained (CG) representation in which some of the atomistic degrees of freedom are removed to reduce the overall model complexity. Specifically, coarse-graining is a molecular modeling technique in which collections of atoms with highly interrelated behavior are reduced to single particles~\cite{Voth2005,Schulten2006,Voth2013,VothandLu2013, Noid2013,Jin2022bottom,Dama2013,VothUCG2014,Brennan2014,Schilling2022noneqCG, cg-marrink,Noid2023, Palma2024}. By neglecting and removing the internal dynamics of the atoms within each group, CG models increase computational efficiency and allow simulations to be performed over increased spatial and temporal scales. 
Coarse-graining has been shown to be capable of reproducing the structural and thermodynamic properties of a broad class of systems including molecular liquids~\cite{Voth2005,craven14b,Moore2014,Noid2015,Moradzadeh2019}, polymers~\cite{Likos2010,craven14d,Wang2013,Ricci2023}, and proteins~\cite{Schulten2006,Voth2013, Noid2013, Rosenberger2021}. 


A limitation of traditional coarse-graining techniques and the corresponding force fields used to perform the CG simulations is transferability -- that is, CG models are optimized at a specific thermodynamic state point, but perform poorly outside of those conditions.~\cite{Rotskoff2023}
The transferability problem arises because the effects of the removed degrees of freedom are themselves a function of thermodynamic conditions.\cite{Noid2016}
Researchers are attempting to tackle the transferability problem in coarse-graining using a variety of techniques, including inverse Monte Carlo methods~\cite{Rosenberger2018}, extended ensemble approaches~\cite{Noid2009, Shell2020} and integral equation methods~\cite{Guenza2018}, among others. A related approach is to use 
machine learning (ML) methods that target thermodynamic consistency~\cite{craven20c,craven20b,Rosenberger2022} to \STRIKE{provide}\ADD{build} data-driven CG models \ADD{that are transferable} across thermodynamic conditions. These CG models \ADD{aim to retain statistical consistency across scales and/or satisfy known relations between thermodynamics variables and their derivative quantities.
Transferability in CG models has been explored, for example, in the context of matching AA configurations across phases, interfaces, and thermodynamic conditions by accounting for both energetic and entropic effects in the CG space \cite{Jin2020transferable,Jin2019entropy, Kidder2021energetic, Pretti2021transferable}.
Developing transferable CG models using ML is currently an open and active research area.
\ADDD{As applications of ML in coarse-graining expand, it is important to assess if CG models developed using data-driven approaches can improve on current methods that seek accurate force field transferability through, for example, the inclusion of correction terms or by fitting to properties other than a single structural observable.
Using ML to mitigate the need to develop CG force fields that rely on quantities averaged across thermodynamic state points is a potential advantage. The flexibility of ML methods can also potentially improve force field generalizability so that a CG model developed on one class of system/molecule can be applied to a different class.}
}

The development of AA and CG force fields is a complex task that involves selection of functional forms, creation and curation of data, optimization of parameters, evaluation of preliminary models, and tuning of hyper-parameters (such as cost functions weights) associated with fitting.\cite{harrison2018review}
Recently, approaches using ML methods have been applied to build many-body models of atomistic~\cite{Kulichenko2021,Ramprasad2017, Wu2023, Unke2021,Smith2017a} and coarse-grained~\cite{cg-nn-thaler, Wang2019machine,Fu2022MLCG,Ricci2023} forces with increased flexibility in comparison to traditional methods. These approaches allow for tight matching with reference data, but introduce even more hyper-parameters to the model and fitting procedure, further complexifying the automation of force field development. \ADD{Machine learning techniques such as neural networks \cite{Husic2020, Loose2023_CGNN, Zhang2018DeePCG, Ruza2020CGionic} and active learning \cite{Duschatko2024uncertainty} 
have been applied to increase the accuracy and reduce the complexity of developing CG force fields. 
Machine leaning applications in coarse-graining include implicit solvation models \cite{Airas2023} and the prediction of solvation energies \cite{Ge2023MLcoarse}.
While including many-body interactions, and effects of those interactions, in CG force fields may improve the accuracy of a CG model, it is more difficult to optimize a complex many-body functional form with simple fitting approaches, e.g. using serial updates to single parameters at a time. Machine learning, however, can provide a computationally tractable approach to quickly fit many-body CG forces when sufficient data is available. The inclusion of many-body forces often results in a better reproduction of the AA configuration distributions \cite{Ge2023MLcoarse, Larini2010threebody, Das2012threemody, Wang2021coarsegrained}.}

\ADDD{One common goal in coarse-graining is to maintain thermodynamic consistency between length scales while at the same time realizing transferability of the CG model across thermodynamic states. 
This multi-objective optimization is often difficult to realize in practice, however,
principally because the multiscale nature of the CG procedure gives rise to nontrivial operations in the construction of force fields and associated mappings between length scales that maintain thermodynamic consistency.
Machine learning models are often able to accurately describe a system's behavior over a wide range of thermodynamic states due to the flexible form of the interpolating functions used in ML, for example, neural networks. 
Advances in ML applied to coarse-graining have been made in parallel with advances in AA force field development using ML-based methods~\cite{Ramprasad2017, Wu2023, Unke2021}. 
However, the application of ML to develop transferable CG force fields is 
currently a complex task with open questions.} 

In this paper, we present an \ourmethod workflow to construct force fields based on the force-matching (or multi-scale coarse-graining) approach\cite{noid2008multiscale1} using the Hierarchically Interacting Particle Neural Network with Tensor Sensitivity (HIP-NN-TS)~\cite{hipnn, hipnn-ts} architecture, which has previously only been applied to AA systems. We show that this workflow is robust in that it is able to consistently build a large number of accurate CG models for a variety of chemical physics systems across many thermodynamic state points \ADD{in a single phase on the phase diagram}. We then use these models to study transferability of the many-body \ourmethod approach. We compare these results to two-body effective potentials using the recently developed OpenMSCG\cite{peng2023openmscg} software. We find that the \ourmethod approach is more consistently accurate, can take advantage of additional training data, and produces more transferable models across varying temperature. We furthermore study several molecules across temperature and density variations, comparing single-state-point models with models produced by training to all available data, and find that both types are surprisingly transferable even as systems undergo large changes in their structural ordering.




%% file: main_sections/methods.tex
\section{Methods}
\label{section:methods}

\begin{figure*}[ht]
    \centering
    \includegraphics[align=c]{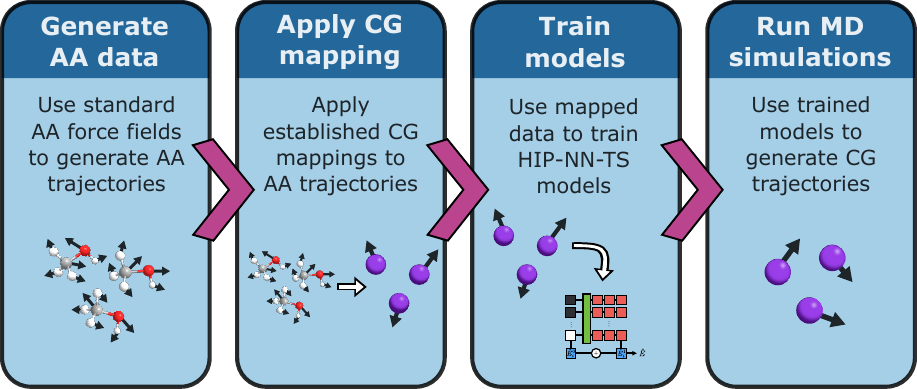}
    \caption{An illustration of the workflow used to create and analyze the \ourmethod models}
    \label{figure: workflow}
\end{figure*}

In this section, we describe training data generation, the ML model architecture, the training procedure, and evaluation methods used in this work. An overview of the workflow is shown in Figure~\ref{figure: workflow}.

\subsection{Coarse-graining} \label{section: coarse-graining}

We use a typical coarse-graining framework in which collections of atoms in an AA system are mapped to CG \emph{beads} and internal degrees of freedom in each collection of atoms are removed from the CG model.
With this aim, a mapping from the AA configuration space to the CG resolution is selected. We use a bottom-up approach to construct the corresponding CG force field, which is defined to preserve the configuration probability density from the AA space in the CG space.\cite{noid2007multiscale}
The relation between the configuration probabilities in each representation is
\begin{equation} \label{equation: CG-AA probability function relation}
    P_{\text{CG}}(\mathbf{R}) = \int P_{\text{AA}}(\mathbf{r}) \delta [\mathcal{M}(\mathbf{r}) - \mathbf{R}] \, d\mathbf{r},
\end{equation}
where $\mathbf{r}=\{\mathbf{r}_1, \mathbf{r}_2, \ldots, \mathbf{r}_n\}$ represents the positions $\mathbf{r}_i$ for atoms $1 \leq i \leq n$ in the AA space, 
$\mathbf{R}=\{\mathbf{R}_1, \mathbf{R}_2, \ldots, \mathbf{R}_{N}\}$ the positions $\mathbf{R}_I$ for CG beads $1\leq I \leq N$, $\mathcal{M}$ the mapping function from $\mathbf{r}$ to $\mathbf{R}$, and $P_{\text{CG}}$ and $P_{\text{AA}}$ the respective CG and AA configuration  probability densities. For the canonical ensemble generated by an AA energy $E_\text{AA}$ at inverse temperature $\beta$, this equation can be cast in terms of a \textit{free energy function} $E_\text{CG}$ as

\begin{equation}
\label{equation:e_cg}
    e^{-\beta E_\text{CG}(\mathbf{R})} \propto \int e^{-\beta E_\text{AA}(\mathbf{r})} \delta [\mathcal{M}(\mathbf{r}) - \mathbf{R}] \, d\mathbf{r}.
\end{equation}
This free energy function is a \emph{potential of mean force} \ADD{(PMF)}, in that the CG forces derived from it should match the force averages across all possible atomistic configurations corresponding to each CG configuration, based on the selected mapping $\mathcal{M}$.
    

Herein each CG bead represents the atoms that compose a single molecule, and we invoke \STRIKE{a}\ADD{the common} center-of-mass mapping to construct the CG representation. \ADD{For a discussion of other popular mapping choices, see Ref.~\citenum{rudzinski2014investigation}.}
We define the mass $M_I$ of a bead $I$ as a sum of the masses $m_i$ for each atom $i$ in the corresponding molecule as
\begin{equation}
    M_I = \sum_{i=1}^{n} m_i \Delta_{i,I},
\end{equation}
where $\Delta$ is an indicator for which atoms correspond to which bead,
\begin{equation}
    \Delta_{i, I} = 
        \begin{cases}
            1, & \text{atom $i$ is in molecule $I$}, \\
            0, & \text{otherwise}.
        \end{cases}
\end{equation}

\STRIKE{The dynamical variable in the coarse-grained model we use is} \ADD{The CG bead coordinates $\mathbf{R}_I$ are calculated from the AA configurations using} the center of mass position, \STRIKE{$\mathbf{R}_I$,} i.e.
\begin{equation}\label{eqn: M}
    \mathbf{R}_I = \mathcal{M}(\mathbf{r})_I = \frac{\sum_{i=1}^{n} m_i \mathbf{r}_i \Delta_{i,I}}{M_I}.
\end{equation}


Also required is a mapping $\mathcal{B}$ from AA forces $\mathbf{f}=\{\mathbf{f}_1, \mathbf{f}_2, \ldots, \mathbf{f}_n\}$ to CG forces  $\mathbf{F} = \{\mathbf{F}_1, \mathbf{F}_2, \ldots, \mathbf{F}_N\}$. Ciccotti, Kapral, and Vanden-Eijnden~\cite{ciccotti2005bluemoon} derived a set of criteria for $\mathcal{B}$ that guarantees that the corresponding free energy function $E_{\text{CG}}$ is consistent with equation~\ref{equation: CG-AA probability function relation}. Summarily, while $\mathcal{B}$ may be nonlinear in $\mathbf{r}$, it must be linear in $\mathbf{f}$, and $\mathcal{B}$ must serve as an inverse of the AA gradient of the coordinate mapping, $\nabla_\mathbf{r} \mathcal{M}$, when contracted over the AA indices. A simple choice consistent with this criteria is to define $\mathcal{B}$ so that $\mathbf{F}_I$ is the sum of forces $\mathbf{f}_{i}$ for atoms $i$ in bead $I$, i.e.
\begin{equation}\label{eqn: B}
    \mathbf{F}_{I} = \mathcal{B}(\mathbf{f})_I = \sum_{i=1}^{n} \mathbf{f}_i \Delta_{i,I}.
\end{equation}
Using these CG mappings, for each frame of data, $\mathbf{F}$ provides an unbiased estimator of the negative derivative of the CG free energy $E_\mathrm{CG}$ with respect to the coordinates $\mathbf{R}$.

\subsection{Model architecture}\label{section: network architecture}

To build the CG free energy function $E_\mathrm{CG}$, we applied the Hierarchically Interacting Particle Neural Network with Tensor Sensitivity (HIP-NN-TS)~\cite{hipnn-ts}.
HIP-NN-TS is a graph-convolutional neural network (GCNN); the convolutions, implemented in an \emph{interaction layer}, make the model invariant under rotations, translations, and permutations of the atoms in the simulation. The tensor-sensitivity component builds upon the original model\cite{hipnn} by introducing many-body features into the individual neurons of the interaction layer. 

We adapted the open-source \texttt{hippynn}\cite{hippynn-repo} codebase that implements HIP-NN-TS to take bead positions as inputs rather than atom positions. The models then predict the free energy of the system based on the bead configurations. The force on each bead is calculated as the negative gradient of the predicted free energy with respect to the bead positions using automatic differentiation\cite{smith2020simple}. Using the language of Ref.~\citenum{hipnn-ts}, the \ourmethod models developed here use tensor order $\ell = 2$ and contain $n_\text{int}=1$ interaction layer, $n_\text{atom}=3$ atomic environment layers, $n_\nu=20$ sensitivity functions, and $n_\text{feature}=32$ atomic features per layer. \ADD{These hyperparameters were determined through trial and error during a preliminary phase of the study based on values which had shown success in previous HIP-NN networks used for atomistic simulations\cite{hipnn, hipnn-ts}. A post-hoc analysis, described further in the Supporting Information section \textit{Neural network hyperparameter investigation}, indicates a good deal of flexibility in selecting these hyperparameters without significantly diminishing the quality of the results.}

In addition to the neural network component of the free energy, a short-ranged pairwise repulsive potential was added to the models. The data used to train the models comes from equilibrium simulations in which there is an effective lower bound on intermolecular distance, $r$. Without the addition of the repulsive potential, the models can generate unphysical, untrained predictions for the forces between pairs of beads whose distance is less than $r$, due to the lack of data. \ADD{Without the repulsive term, the GCNN potential would have significant uncertainty in the small-$r$ region. This is because those configurations are not seen in the training data and, without the replusive term, small inter-particle distances would constitute significant extrapolation for the potential. 
By including the repulsive term, we add a physics-based term (in contrast to the ML-based component) that alleviates the sampling problem. \ADDD{This is a common technique to employ in such a situation, although the exact forms of the short-range potentials vary.\cite{husic2020coarse, fellman2024fast, byggmastar2019machine}}
The repulsive potential improves the stability of the model because extrapolation of the potential to small inter-bead distances is no longer performed.}
The repulsive potential ensures that any two beads separated by less than $r$ are repelled.\STRIKE{, as implied by the lack of data for that region of phase space.}
The repulsive potential is of the form 
\begin{equation}
    E_\text{rep}(r) = E_0 e^{-ad}
\end{equation}
where $r$ is the distance between the pair of beads, and $E_0, a > 0$ are parameters set based on the specific system. Importantly, because the pairwise potential pertains to lack of data, it is necessary to set $E_0$ and $a$ before training the neural network. The procedure used for identifying $E_0$ and $a$ is detailed in the Supporting Information section \textit{Repulsive potential parameterization details}. We emphasize that the repulsive pair potential significantly improves the stability of learned \ourmethod models during simulations, resulting in a highly automated workflow. 

\ADD{Example scripts for training such a model and for using the resulting model to run coarse-grained MD are included in the open-source \texttt{hippynn}\cite{hippynn-repo} repository.}

\subsection{Training}
\label{section: training}

For the methanol comparison study, we follow the data generation scheme described in an OpenMSCG tutorial~\cite{open-mscg-tutorial} for building CG models of methanol. The molecular dynamics suite GROMACS~\cite{lindahl2001gromacs} is used to generate 100,000 timesteps of 1 fs each of a box with 1,728 methanol molecules under periodic boundary conditions. Every 100th step is saved, resulting in 1,000 AA frames. The topology and initial coordinates are downloaded from the GROMACS webpage and the OPLS-AA~\cite{OPLSAA} force field is used. The simulation is run in the canonical ensemble using a Nos\'e-Hoover thermostat.


For the cross-molecular study, we use the LAMMPS~\cite{lammps} software package with the \ADD{all-atom (i.e. non-constrained)} GROMOS-54A7~\cite{gromos-54A7} force field to simulate three molecular fluids: methanol, benzene, and methane across a variety of temperatures and densities above the critical temperature for each molecule. The molecular topologies were obtained using the Automated Topology Builder (ATB) and Repository~\cite{atb1, atb3}. PACKMOL~\cite{packmol} was used to generate initial coordinates and Moltemplate~\cite{moltemplate} was used to generate \STRIKE{PACKMOL}\ADD{LAMMPS} input files. We performed MD simulations in the canonical ensemble using a Nos\'e-Hoover thermostat. Each simulation contained 1,024 molecules and the density was controlled using cubic boxes of various size with periodic boundary conditions. Following an equilibration procedure, each MD simulation consisted of 50,000 timesteps of 1 fs. Every 50$^{\text{th}}$ frame was recorded, totalling 1000 frames.

\begin{figure*}[ht]
    \centering
    \includegraphics[align=c]{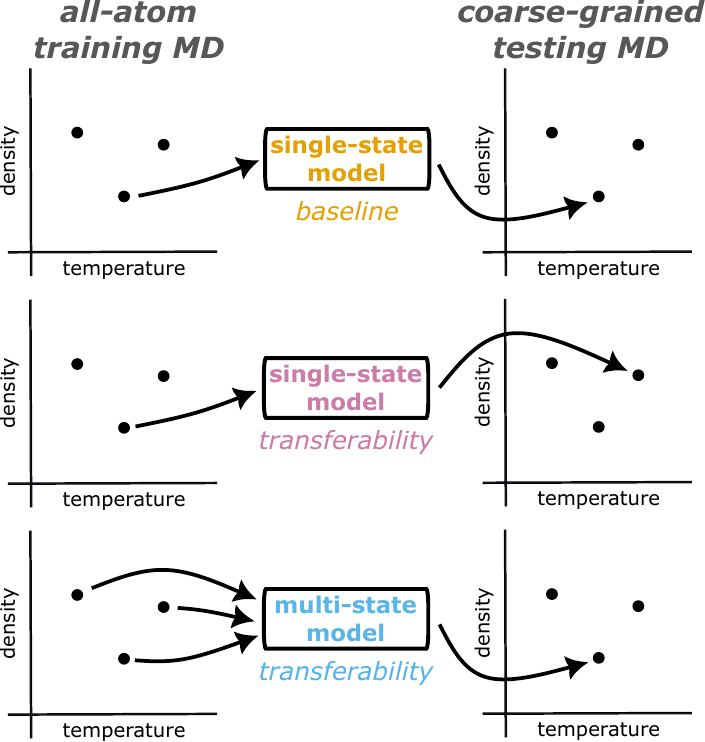}
    \caption{An illustration of the three types of tests performed on the \ourmethod method.}
    \label{figure: model test types}
\end{figure*}

For both studies, after generating the AA data, the mappings $\mathcal{M}$ of Eq.~\ref{eqn: M} and $\mathcal{B}$ of Eq.~\ref{eqn: B} were applied to each frame to create the training data for the \ourmethod models. For the methanol comparison study and the cross-molecular study, we trained \emph{single-state} models using data from a single state point (i.e. for each temperature and density combination). These single-state models are then applied using MD at the same state point to which they were trained to establish a baseline. They are also applied with MD at other temperatures and densities, to test their transferability. For the cross-molecular study, we also trained a \emph{multi-state} model for each molecule using combined data from every state point and tested it using MD at each of those state points. See Figure~\ref{figure: model test types} for an illustration of these three different test schemes.

The multi-state datasets are the same size as the single-state datasets, built by evenly subselecting the frames from each state point. In each case, the 1000 frames of training data are split randomly into 800 training frames, 100 validation frames, and 100 testing frames. 
The loss function used to train the \ourmethod models was the sum of the root mean square error (RMSE) and mean absolute error (MAE) for the model predictions versus the training data values of the forces for each configuration in the dataset. \ADD{Both the MAE and RMSE were included in the loss function because the combination was found to be successful in prior works\cite{hipnn, hipnn-ts}. A post-hoc investigation, described i Supporting Information section \textit{Neural network hyperparameter investigation}, suggests that using RMSE but not MAE (as prescribed by the MSCG formalism and corresponding PMF) to train a model has minimal impact on the radial distribution function (RDF) and angular distribution function (ADF) error presented later.}

\subsection{Performance evaluation}

We measure the accuracy of CG models by how well they reproduce statistics from the AA models, in accordance with Eq.~\ref{equation: CG-AA probability function relation}.
In particular, the
\STRIKE{radial distribution functions (RDFs)} 
\ADD{RDFs} 
measured during MD simulations using the CG models were compared to the RDFs generated from the AA MD simulations.
The RDF, denoted by $g(r)$,  describes the statistics for finding a particle at distance $r$ from a randomly chosen particle. It is normalized so that an RDF is asymptotically one for large radii by defining the RDF as the ratio between the local density variations and the bulk density.~\cite{Rosenberger2020}
The RDF is a key structural metric used in characterizing the degree of local ordering in a system. It can be determined using theoretical, computational, or experimental approaches, and can be used to derive much thermodynamic information about a system~\cite{Trokhymchuk2005,hansen06,GrayG1984,Foidl1986exact,Rosenberger2020}. In this work, AA RDFs are computed using molecular centers of mass, and CG RDFs, correspondingly, are computed using the bead positions. \ADD{As a ratio between densities, the RDF has no units; we denote this as (-) in the figures.}

In order to construct the CG RDFs, we performed CG MD simulations in the canonical ensemble. The initial positions for these simulations were taken from a random frame of the model training data at the appropriate temperature and density. After equilibrating the CG system, 50,000 timesteps of 1 fs each were run, and each 50$^{\text{th}}$ frame  was recorded. The RDFs were computed from these 1,000 frames of data. The AA RDFs were taken from the AA training data, which was also performed using canonical ensemble MD as detailed in the \nameref{section: training} subsection.

To quantify the difference between the AA and CG RDFs, we use the \emph{total absolute error} (TAE). The TAE is the total area between two curves, i.e.,
\begin{equation}
    \text{TAE}(g_1, g_2) = \int_0^\infty |g_2 - g_1| \, dr.
\end{equation}
For the calculations in this paper, we used a finite sum approximation of TAE. Specifically, we used
\begin{equation}
     \text{TAE}(g_1, g_2) \approx \sum_{i=0}^{n} | g_2(r_{i}) - g_1(r_{i}) | \Delta_{r},
\end{equation}
where $\{r_j\}$ represents a sequence of distance values, evenly spaced between $r_0 = 0$ and an appropriately chosen $r_n = r_{\text{max}} > 0$, with width $\Delta_{r}$ between each pair of successive values. We chose \STRIKE{TAE to compare RDFs rather than MAE} \ADD{to compute the TAE between RDFs rather than the MAE between RDFs} because the \ADD{RDF} TAE is not strongly sensitive to the choice of upper cutoff $r_\text{max}$ of the RDF radius, whereas the \ADD{RDF} MAE is.

\ADD{Additionally, we compare the ADFs from the \ourmethod MD runs to those from the AA MD simulations. The ADF of a trajectory describes the distribution of angles between triples of points in each frame of a trajectory. This higher-order structural function provides further insight into the arrangements of particles in the trajectory beyond what is contained in the RDF. More precisely, the ADF is the probability distribution of angles in the range $[0^\circ, 180^\circ]$ where each angle is formed by a central particle to two neighbors, where both arms of the angle from the central particle have length less than a set cutoff distance $r_\text{max}$. In general, higher values of $r_\text{max}$ will produce less structured ADFs, as the behavior of particles at greater distances from one another is less correlated. As with the RDF, the ADF is a ratio, and hence has no units, which we denote as (-) in the figures.
Unlike the RDF, for the ADF, either TAE or MAE can be appropriate to quantify the difference between two histograms, as the range of values for which the distribution is defined is fixed at $[0^\circ, 180^\circ]$. We use ADF MAE here, again approximated by a finite sum, which is dimensionless.} 

A usual method for evaluating ML models is \STRIKE{to examine error metrics such as MAE and RMSE on test data reserved from training.} \ADD{to compare predicted model outputs (forces, in our case) with the corresponding true values from test data withheld during training.} However, for the CG problem here, the training data $\mathbf{F}$ is a statistical distribution of forces, whereas the model predicts the mean of this distribution, $\langle \mathbf{F} \rangle$; as a result, there is intrinsic noise in the loss function.\cite{Kramer2023} For this reason, conventional metrics are not easily usable for assessing the quality of the \ourmethod model. Nevertheless, several conventional metrics for the performance of the models are reported in the Supporting Information\ADD{ section \textit{Training statistics}.}
\STRIKE{, and we do observe that within a given problem, there is some correspondence between the error given by RDF TAE and by force MAE/RMSE.}

%% file: main_sections/results.tex
\section{Results}
\label{section:results}

In this section, the performance of the \ourmethod models is explored. Both \emph{single-state} models, those trained using data at a specific temperature and density, and \emph{multi-state} models, those trained using data from a range of temperatures and densities, are discussed. 


\subsection{Methanol Comparison Study}\label{subsection: methanol comparison study}

\begin{figure*}
    \includegraphics[align=c]{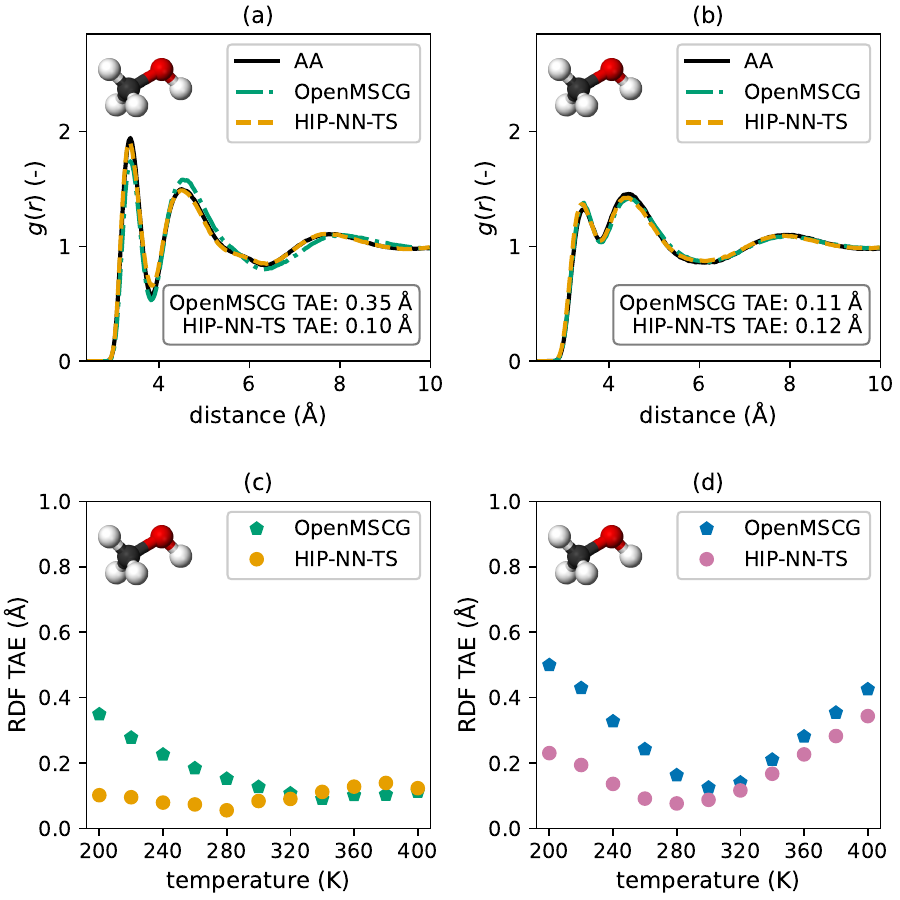}
    \caption{Subfigures (a) and (b) show a comparison of methanol RDFs generated using (1) a reference AA simulation, (2) the MS-CG technique, and (3) the single-state \ourmethod models. These RDFs were generated at (a) 200 K, density 0.77 g/cm$^3$ and (b) 400 K, density 0.77 g/cm$^3$. Subfigure (c) summarizes the corresponding results across 11 temperatures. Subfigure (d) shows the transferability of a single-state model for each method, each trained with 300 K data.
    }
    \label{figure: rdfs, mscg}
\end{figure*}

To establish a baseline for the \ourmethod approach, we compared it to the recently released OpenMSCG software for coarse-graining methanol.~\cite{peng2023openmscg}. OpenMSCG is based upon methods proposed by Izvekov and Voth~\cite{izvekov2005multiscale1,Voth2005} and later further developed and generalized by others including Noid et.~al.\cite{noid2007multiscale,noid2008multiscale1,noid2008multiscale2}. The OpenMSCG software provides a set of \emph{force-matching} routines which implement a bottom-up coarse-graining method that calculates the effective CG interactions by minimizing the difference between CG forces and reference AA forces; this is quite similar to our workflow and uses the same coordinate and force mappings. The main difference is that OpenMSCG uses a pairwise force/energy function between beads instead of a GCNN. 

The result\STRIKE{s}\ADD{ing RDFs for}\STRIKE{ of} this comparison are given in Fig.~\ref{figure: rdfs, mscg}. Both methods perform well, with the \ourmethod method exhibiting significantly lower error at lower temperatures as shown in Figure~\ref{figure: rdfs, mscg}(a) for 200 K. These results illustrate that, in general, we expect the neural network-based methodology to result in CG force fields that perform as well as or better than those constructed using force-matching when applied to data at a specific state point. 
Figure~\ref{figure: rdfs, mscg}(b) shows results for methanol at 400 K. In this case, the OpenMSCG and \ourmethod methods give very similar and excellent overall accuracy. As such, in this case, there is only a limited potential for improvement over the OpenMSCG model.
We performed this study for a range of temperatures from 200 K to 400 K, with the overall RDF TAEs presented in Figure~\ref{figure: rdfs, mscg}(c). At lower temperatures, the \ourmethod models perform substantially better than the OpenMSCG models. As the the temperature is increased, the difference between the two methods is less pronounced. In the temperature range 340 K-380 K, the OpenMSCG method is slightly more accurate, although the difference between the methods is relatively small. The models shown in Figure~\ref{figure: rdfs, mscg}(c) were trained using 1000 frames of training data. In the Supporting Information Figure~S2, we show similar results constructed using 10 and 100 frames of training data. With far fewer data, the advantage of \ourmethod over OpenMSCG is less pronounced; we find the performance of OpenMSCG saturates more quickly as the dataset size is increased.

\begin{figure*}
    \includegraphics[align=c]{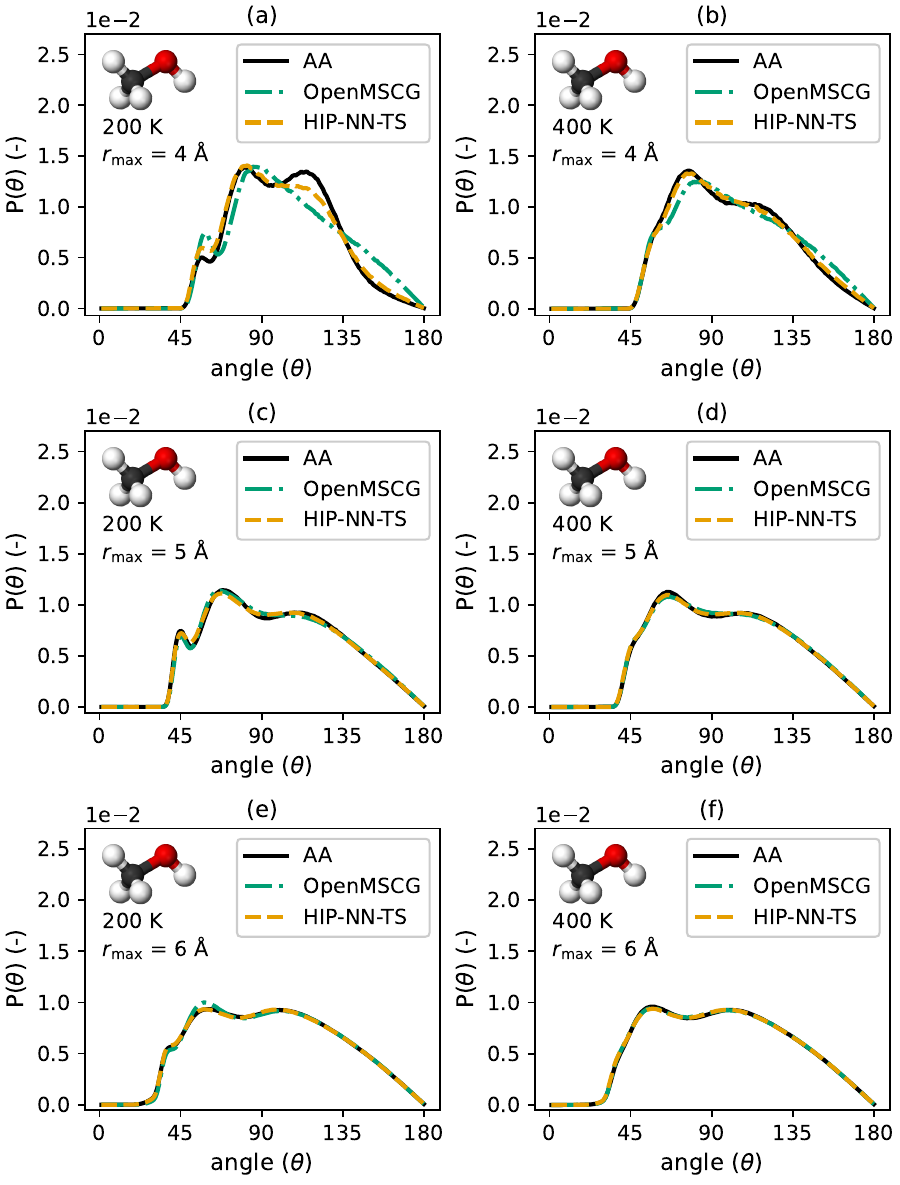}
    \caption{\ADD{Comparison of methanol ADFs generated using (1) a reference AA simulation, (2) the MS-CG technique, and (3) the single-state \ourmethod models. Subfigures (a), (c), and (e) show ADFs for 200 K, density 0.77 g/cm$^3$ with ADF cutoff values $r_\text{max}$ of 4~\AA, 5~\AA, and 6~\AA\ respectively. Subfigures (b), (d), and (f) show ADFs for 400 K, density 0.77 g/cm$^3$ with the same respective cutoff values.
    }}
    \label{figure: adfs specific, mscg}
\end{figure*}

\begin{figure*}
    \includegraphics[align=c]{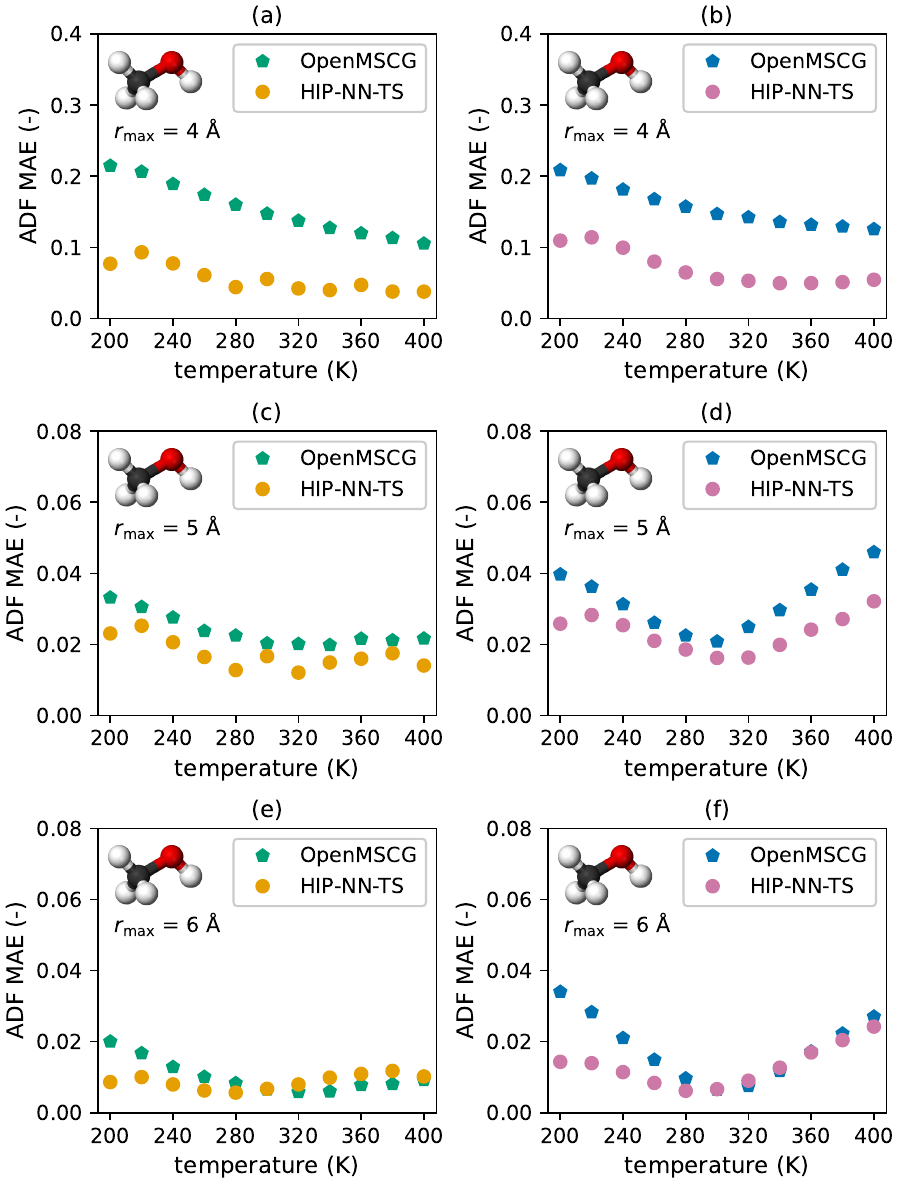}
    \caption{\ADD{Summary of ADF comparisons for the MS-CG technique and the \ourmethod technique against a reference AA simulation. Subfigures (a), (c), and (e) show results for the single-state baseline models using cutoff value $r_\text{max}$ of (a) 4~\AA, (c) 5~\AA, and (e) 6~\AA. Subfigures (b), (d), and (f) show results for the single-state transferability test using the 300 K model for each method with ADF cutoff values $r_\text{max}$ of (b) 4~\AA, (d) 5~\AA, and (f) 6~\AA. 
    }
    }
    \label{figure: adfs summary, mscg}
\end{figure*}

\ADD{When comparing the ADFs generated with the OpenMSCG method and the \ourmethod method against the AA reference data, shown in Figure~\ref{figure: adfs specific, mscg} and in Figure~\ref{figure: adfs summary, mscg} parts (a), (c), and (e), we see again that the \ourmethod models perform better or similarly to the OpenMSCG models in every case. The difference is most stark for the ADFs generated with cutoff value $r_\text{max} =$ 4~\AA, where the ADFs are most structured. The MAE between the OpenMSCG ADFs with the reference AA ADFs is around twice that for the \ourmethod models at every temperature tested. When the ADF cutoff value $r_\text{max}$ is increased to 5~\AA\ or 6~\AA, this difference largely disappears and both methods reproduce the ADFs very closely. 
}

To understand the transferability of the models, we applied the single-state model learned at 300 K for each method to a range of temperatures and recorded the TAEs of the resulting RDFs in Figure~\ref{figure: rdfs, mscg}(d). In both cases, the model produces lower RDF errors in the vicinity of the training state point, and the TAE rises smoothly at higher and lower temperatures. However, the
\ourmethod model shows significant lower RDF TAE compared to OpenMSCG when applied at temperatures further from the training state point. At temperatures lower than 300 K, the TAE for the \ourmethod models is approximately a factor 2 lower in comparison to OpenMSCG. \ADD{The corresponding ADF errors are show in Figure~\ref{figure: adfs summary, mscg} parts (b), (d), and (f). For cutoff value $r_\text{max}=$ 4~\AA\ (subfigure (b)), the MAEs for the \ourmethod ADF is approximately half the MAE for the OpenMSCG model at each temperature. For cutoff value $r_\text{max}=$ 5~\AA\ (subfigure (d)), this difference is lessened, but the \ourmethod model still slightly outperforms the OpenMSCG model in every case. Finally, with cutoff value $r_\text{max}=$ 6~\AA\ (subfigure (f)), the MAE error values for the OpenMSCG ADFs are nearly identical to the \ourmethod error values for temperatures between 300 K and 400 K, while the OpenMSCG errors are closer to double that of the \ourmethod models at temperatures from 200 K to 280 K. 
}

\STRIKE{As the main difference in the methods is the many-body nature of the \ourmethod model vs. the pairwise nature of OpenMSCG, these} 
\ADD{The primary difference between the \ourmethod method and the OpenMSCG method is that the \ourmethod models incorporate many-body effects while the OpenMSCG methods are pairwise. These}
results illustrate how accounting for many-body interactions may be relevant to producing more transferable CG force fields. 

\subsection{\STRIKE{Cross-molecular Study}\ADD{Multi-molecule Study}}\label{subsection: multi-molecule}
\label{sec:multimol}

To further study thermodynamic transferability in \ourmethod models, we applied the \ourmethod method to several molecules (methanol, benzene, and methane) across a range of temperatures and densities as described in the \nameref{section: training} subsection. 

\begin{figure*}[t]
    \includegraphics[align=c]{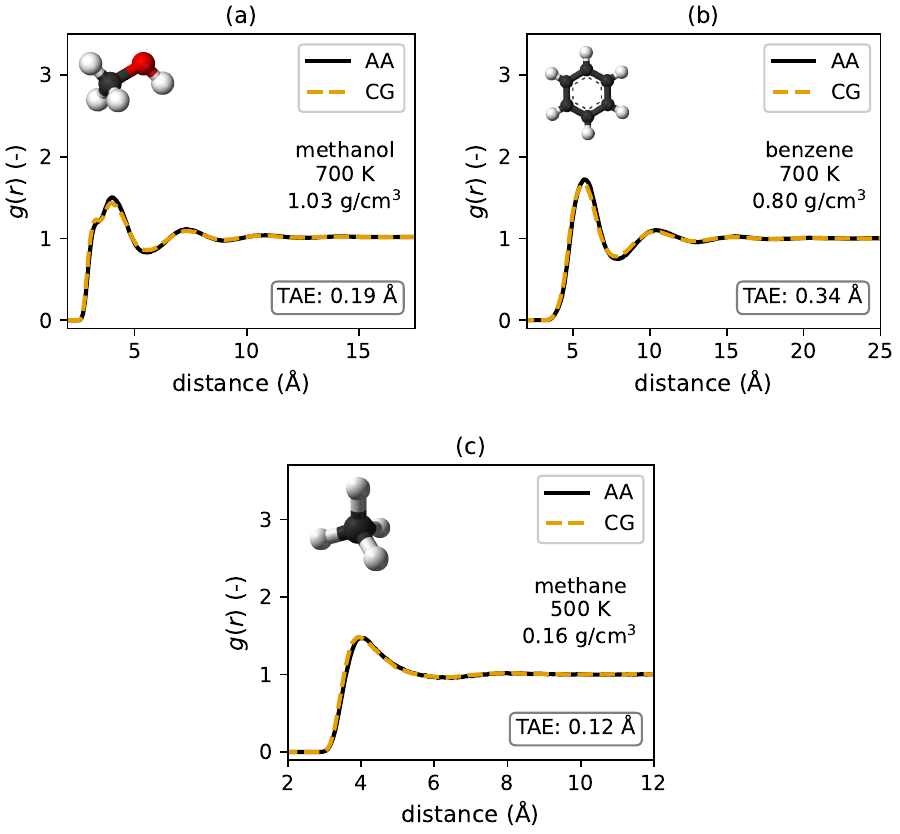}
    \caption{Baseline comparison of reference RDFs generated from AA MD against RDFs generated via a \ourmethod single-state model for (a) methanol, (b) benzene, and (c) methane.}
    \label{figure: rdfs, specific}
\end{figure*}

Figure~\ref{figure: rdfs, specific} shows examples of the RDFs generated using single-state models for methanol, benzene, and methane. Each subfigure shows a reference AA RDF computed using molecular centers-of-mass for atomistic MD data generated at the same temperature and density. There is excellent agreement between the AA and CG RDFs in each case, illustrating that the \ourmethod models accurately capture the shape and magnitude of the RDF peaks for each of the examined molecular fluids.

\begin{figure*}
    \includegraphics[align=c]{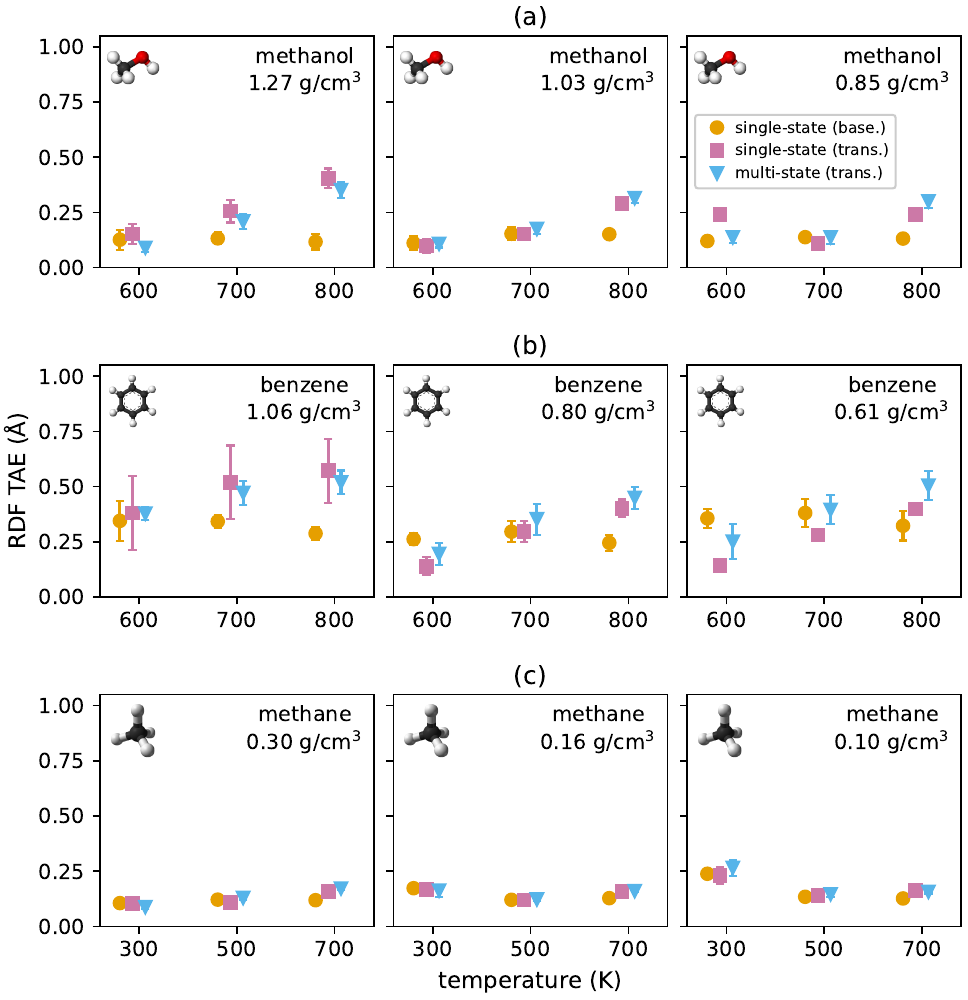}
    \caption{Total absolute error (TAE) between RDFs generated by AA MD and using various \ourmethod models for (a) methanol, (b) benzene, and (c) methane. The horizontal placement of the markers has been offset slightly for visual clarity. The error bars show the standard deviation calculated by constructing five \ourmethod models randomly sampled as  described in the text. 
    }
    \label{figure: rdf taes summary}
\end{figure*}

Figure~\ref{figure: rdf taes summary} shows the results of three workflows for all three molecules studied across the nine state points used for each molecule. First, the single-state models are used to provide a baseline of a usual CG approach which trains and tests at the same thermodynamic state point (denoted \emph{single-state (base.)} in the figure). Second, for each molecule, the single-state model trained at the center state point is applied to each of the test state points (denoted \emph{single-state (trans.)}), providing a picture of the transferability of the \ourmethod model when extrapolating through thermodynamic state space. Finally, the multi-state model is applied to each state point (denoted \emph{multi-state (trans.)}) to test whether training to all state points provides improved transferability characteristics (for details see the \nameref{section: training} section). 
Each molecular fluid was studied at three densities (shown in three panels) and three temperatures (shown on the vertical axis of each panel) in Figure~\ref{figure: rdf taes summary}. 
The error bars show the standard deviation calculated across five trials, where in each trial, several factors were randomly varied: the train/valid/test split for the data used to create the \ourmethod model, the initial weights of the HIP-NN-TS network, the initial frame for the \ourmethod dynamics, and the random number seed used for the thermostat. \ADD{Of the 405 trajectory comparisions used to create this figure, some of those with the highest TAEs are shown in Supporting Information Figure~S5.}

The findings for the single-state baseline test were consistent for the three molecules. For methanol (Figure~\ref{figure: rdf taes summary}(a)), we observe that the single-state models generate similar, very low ($<0.25$~\AA) TAE values over all the densities and temperatures studied. 
For benzene (Figure~\ref{figure: rdf taes summary}(b)), the average TAE is good ($<0.45$ ), although somewhat higher than methanol, but again, single-state models produce fairly consistent TAE values across all state points.
For methane (Figure~\ref{figure: rdf taes summary}(c)), the single-state models perform best at the lowest temperature and highest density studied. This is an interesting observation because at this state point we expect the RDF to have the most structure and stronger correlations at larger $r $ values in comparison to, for example, higher temperatures and lower densities.
For the high density state points, the \ourmethod model for methane performs worse as the temperature is increased. Overall, the results in Figure~\ref{figure: rdfs, specific} and Figure~\ref{figure: rdf taes summary} illustrate that the \ourmethod methodology developed in this article can be applied to construct CG free energy functions that generate RDFs in strong agreement with AA results. 

For the transferability test of the single-state models, the results show a surprising level of transferability for \ourmethod models; in general the TAEs for this test are almost always within a factor of 2 of the baseline single-state performance, with the exception of high-temperature, high-density methanol. Even more surprising, for several state points the single-state transferability performance is actually superior to the baseline models. Another observation is that significant variance between runs is observable for high-density benzene. Under more careful examination, these models produced slightly under-structured RDFs, and the variance between models is explained by the degree of under-structuring. A similar but less pronounced effect is visible for high-density methanol, where some variance appears for the single-state transferability model. In this case, the RDFs were not definitively under- or over-structured. Despite these fluctuations, the RDF quality is still reasonable.




\begin{figure*}
    \includegraphics[align=c]{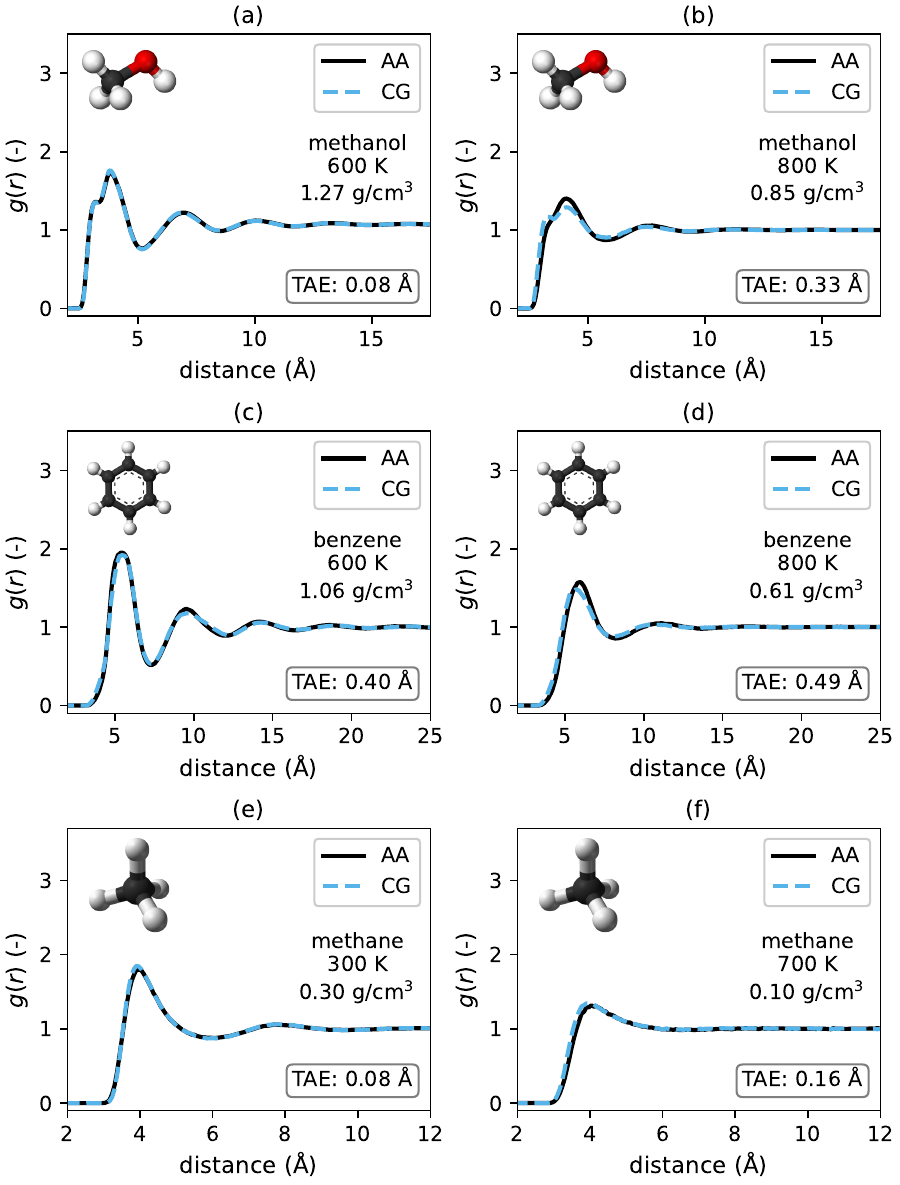}
    \caption{Comparison of reference RDFs generated by AA MD against RDFs generated via the \ourmethod multi-state model for (a) and (b) methanol, (c) and (d) benzene, and (e) and (f) methane. The state points represented are the left-most and right-most pictured in Figure \ref{figure: rdf taes summary} for each molecule, which have, respectively, the most and least structured RDFs.}
    \label{figure: rdfs, multi-state}
\end{figure*}

Like the single-state transferability tests, the multi-state models exhibit remarkable transferability, in some instances even performing better than the single-state models. The higher variance observed in the single-state transferability tests for benzene has been significantly reduced.
More detailed investigation of the underlying RDFs demonstrates that the multi-state \ourmethod models reproduce the AA statistics even through relatively large changes in the structural ordering of the fluid. Fig.~\ref{figure: rdfs, multi-state} shows the most extreme state points (high-density, low-temperature and low-density, high-temperature) with respect to structural ordering. All of the molecules undergo significant changes in structural ordering across the space of thermodynamic states. We also tested whether the multi-state model can be productively applied to state points not present in the training data.  We examined this for methanol by generating AA ground truth at two intermediate densities and two intermediate temperatures, yielding four new state points.  The results of the multi-state model at these previously unseen points are shown in Supporting Information Figure~S4, and demonstrate that the performance of the model does not significantly change when it is applied to intermediate state points not seen during training.

\begin{figure*}
    \includegraphics[align=c]{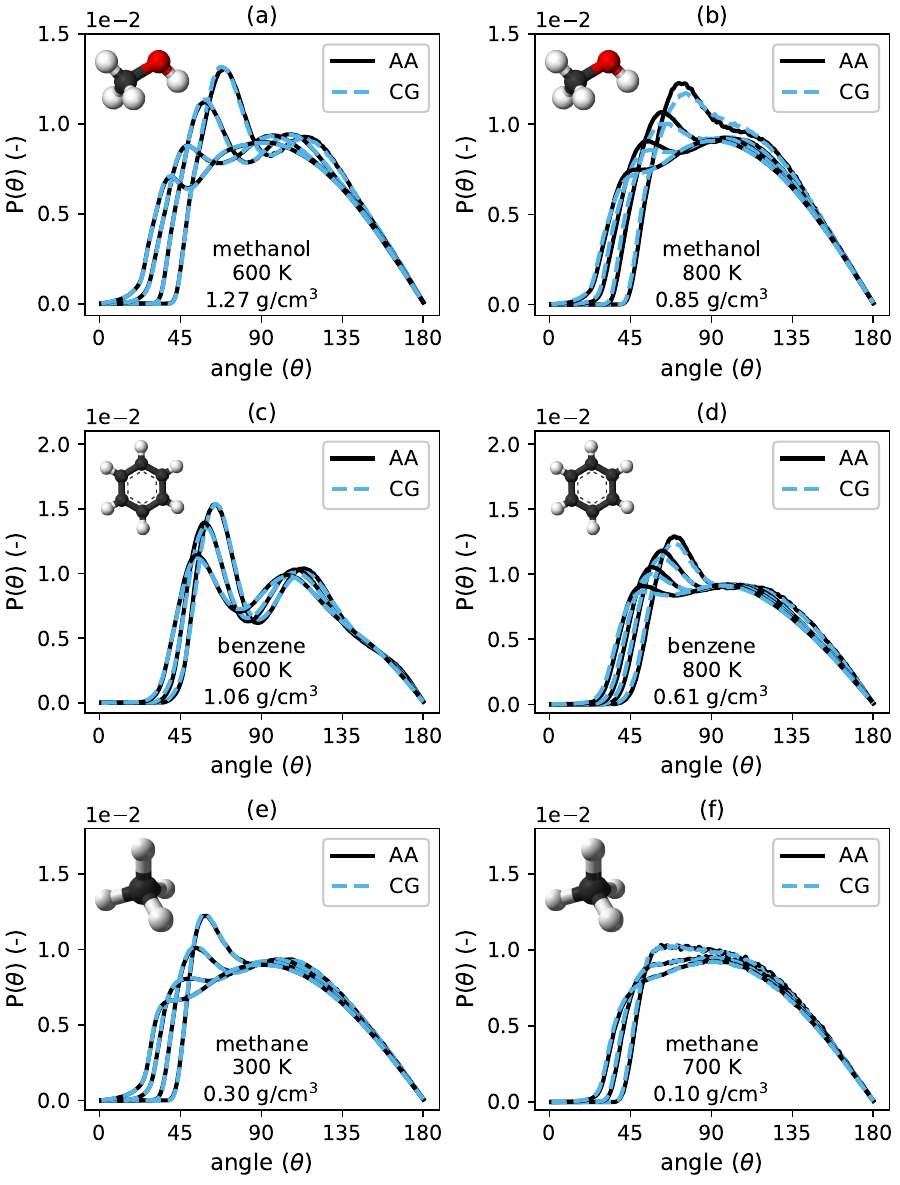}
    \caption{\ADD{Comparison of reference ADFs generated by AA MD against ADFs generated via the \ourmethod multi-state model at several different cutoff values $r_\text{max}$ for (a) and (b) methanol, (c) and (d) benzene, and (e) and (f) methane. 
    The cutoff values used are 4~\AA, 5~\AA, 6~\AA, and 7~\AA\ for methanol; 6~\AA, 7~\AA, 8~\AA, and 9~\AA\ for benzene; and 5~\AA, 6~\AA, 7~\AA, and 8~\AA\ for methane. The smaller cutoff values in each case correspond to the more structured ADFs.}}
    \label{figure: adfs, multi-state}
\end{figure*}

\begin{figure*}
    \includegraphics[align=c]{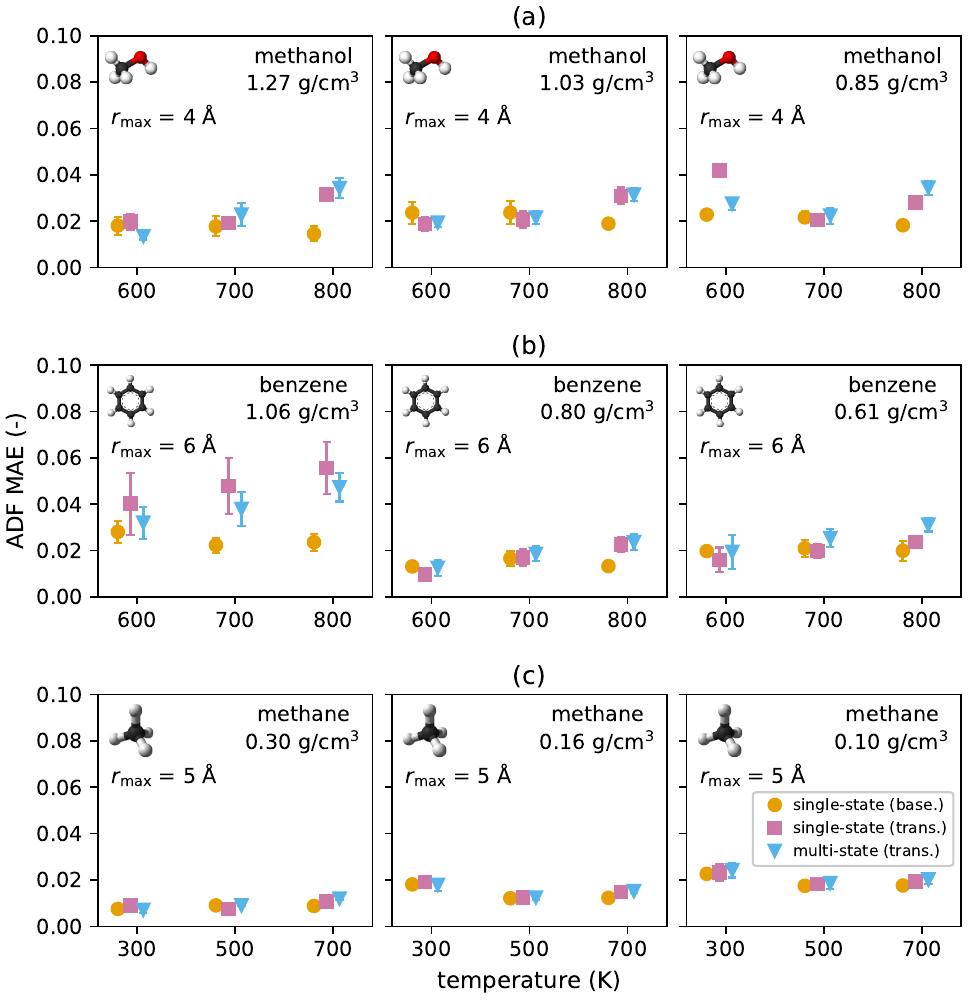}
    \caption{\ADD{Mean absolute error (MAE) between ADFs generated by AA MD and using various \ourmethod models for (a) methanol (with cutoff value of $r_\text{max}=$ 4~\AA), (b) benzene (with cutoff value of $r_\text{max}=$ 6~\AA), and (c) methane (with cutoff value of $r_\text{max}=$ 5~\AA). The horizontal placement of the markers has been offset slightly for visual clarity. The error bars show the standard deviation calculated by constructing five \ourmethod models randomly sampled as described in the text.}}
    \label{figure: adfs, summary}
\end{figure*}

\ADD{
Further, when comparing ADFs generated using AA MD against the ADFs generated with the \ourmethod method, we see strikingly similar results. ADFs from selected multi-state transferability tests are shown in Figure~\ref{figure: adfs, multi-state}, with the results pictured in each subfigure coming from the same trial as the results in the corresponding subfigure in Figure~\ref{figure: rdfs, multi-state}. Each subfigure shows the ADFs corresponding to several choices of ADF cutoff value $r_\text{max}$. The only pictured case in which there is visible deviation from the reference ADFs is for methanol at 800 K with density 0.85 g/cm$^3$, where the ADFs for cutoff values $r_\text{max}$ of 4~\AA, 5~\AA, and 6~\AA\ exhibit slightly less structure than the corresponding reference ADFs. The ADF results for all molecules, tests, and state points at selected cutoff values are displayed in Figure~\ref{figure: adfs, summary}. 
Several of the ADFs with the highest MAE values among those captured in this figure are provided in Supporting Information Figure S6.
Many similarities to the RDF results shown in Figure~\ref{figure: rdf taes summary} can be found. For each molecule, the single-state baseline test results were largely consistent for each molecule. For methanol (Figure~\ref{figure: adfs, summary}(a)), the ADF MAE values are around 0.02 for each temperature and density combination. For benzene (Figure~\ref{figure: adfs, summary}(b)), the ADF MAEs are again all near 0.02, with slightly more fluctuation between state points than what was observed for methanol. Finally, for methane (Figure~\ref{figure: adfs, summary}(c)), the ADF MAEs range from around 0.01 to 0.02, with improved performance as the density increases. These results provide further evidence that the single-state \ourmethod models are able to closely replicate the structure of the AA trajectories used to train them. 
}

\ADD{Figure~\ref{figure: adfs, summary} also shows the ADF MAEs for the transferability test of the single-state models. Again, the results further support the claim of transferability of these single-state \ourmethod models. In many instances, the ADF error for the transferability test model is nearly equivalent to the ADF error for the single-state baseline model. The largest increases in ADF error when performing the transferability test were for benzene at 1.06 g/cm$^3$, the highest density tested, followed by methanol at 800 K and 1.27 g/cm$^3$ and at 600 K and 0.85 g/cm$^3$. In all cases, the ADF MAE is no more than three times the baseline result. 
}

\ADD{Finally, Figure~\ref{figure: adfs, summary} includes the results for the multi-state transferability test ADFs. We observe very similar results for the multi-state and single-state transferability tests in almost every instance, with the highest error for the multi-state transferability test ADFs being for benzene at 1.06 g/cm$^3$, followed by several temperature and density combinations for methanol. Again, in each instance the ADF MAE is no more than three times the baseline result. And just as observed with the RDF errors, there is fluctuation between whether the single-state or multi-state transferability test results are better. This finding reinforces the claim of transferability of both the single-state and multi-state \ourmethod models across temperature and density. 
}

The \ourmethod workflow produces accurate models for a range of molecules and state points without tuning. The main parameter that might be adjusted is the interaction radius of the HIP-NN-TS architecture, but in this work, a uniform interaction distance of $12$~\AA~was used for all molecules.

\subsection{Computational cost}

Computation times for \ourmethod training and MD are affordable enough to enable high-throughput analysis. \ADD{In the multi-molecule study,} for benzene and methane, training the \ourmethod models and running MD with the trained model each consistently took less than 30 minutes with a single NVIDIA A100 GPU. For methanol, each took consistently under one hour with the same architecture. As a result, in the course of this work, we were able train more than 150 models and run more than 400 MD simulations. Scientifically, the most important aspect of automating these simulations was the addition of the repulsive potential determined as a pre-training step.

\ADD{
For the results in the \nameref{subsection: methanol comparison study} subsection above, the \ourmethod method takes about 45 minutes to run 100,000 steps on 1,728 beads, whereas the pairwise OpenMSCG method takes a few minutes to run 100,000 time steps. Given the extremely optimized implementation of pairwise tabular potentials in LAMMPS, a factor of about 20 between the OpenMSCG example and \ourmethod potential is better than expected. 
}

We also performed a comparison of our \ourmethod models to the AA HIP-NN-TS models of Ref.~\citenum{hipnn-ts} using a periodic box of 1,024 methanol molecules. The AA ML potential could compute the forces approximately 0.3 s on average, where the CG potential could compute the forces in approximately 0.04 s on average. There is also a factor of four between the typical timestep for AA MD with an ML potential (0.25 fs) and the timestep used here (1 fs), leading to an overall 30 times faster simulation capacity with the \ourmethod method.

\ADD{ All coarse-grained MD simulations in this manuscript were performed using a timestep of 1~fs. A timestep of between 0.5~fs and 2~fs is common for coarse-grained neural network models. For example, Ref.~\citenum{durumeric2024learning} uses 2~fs and Ref.~\citenum{loose2023coarse} uses 2~fs and 0.5~fs depending on the model. We tested a timestep of 5~fs for one MD run of each type (single-state baseline, single-state transferability test, multi-state transferability test) for each of the nine methanol temperature/density combinations used in the \nameref{subsection: multi-molecule} subsection. We tested energy conservation using this 5 fs timestep, which was satisfactory in 17 of the 27 models. The models that failed energy conservation were mostly for high-density or high-temperature configurations. Improving the dynamics of the \ourmethod models could increase their ability to conserve energy with larger timesteps at high densities and temperatures, allowing for an even greater speed-up when compared with AA CG methods.  Overall this indicates that our overall speedup factor over all-atom HIP-NN-TS can thus be estimated as in the range 30x to 150x.}

Of course, there are many \ADD{other} factors which will affect this number, \STRIKE{such as variations in the timesteps used,}
variations in the network architecture hyperparameters, the overall number of particles considered (1024 particles in the CG simulation does not typically saturate the GPU) and perhaps most importantly, variations in the number of atoms which are coarse-grained over. In regards to this last factor, methanol has a 6-to-1 ratio between atomistic and CG representations, but benzene has a 12-to-1 particle ratio between atomistic and CG representations, and furthermore has fewer neighbors in the same interaction radius, leading to reduced cost for message-passing operations in the neural network.





%% file: main_sections/conclusion.tex
\section{Conclusions}
\label{section:conclusion}

In this work, we built an \ourmethod workflow based on the bottom-up approach of Multi-Scale Coarse-Graining, for the first time using the Hierarchically Interacting Particle Neural Network (HIP-NN) \cite{hipnn,hipnn-ts} with the aim of studying the transferability of coarse-grained models through varying thermodynamic conditions. In order to produce a robust workflow suitable for studying many molecules and state points, it was important to include a pairwise-repulsive potential, with parameters set before training. Otherwise, the workflow is very similar to training an AA potential using (fluctuating) forces, but not energies. As a result we were able to build more than 150 models and run over 400 MD simulations with those models, exploring the accuracy of models trained and used at the same state point, models trained and used at different state points, and models trained and used at multiple state points.

Our results show that the \ourmethod models here produce more consistent structural accuracy (as quantified by the RDF \ADD{and ADF}) than OpenMSCG pairwise CG models built using the same data, using liquid methanol at a variety of temperatures. Furthermore, training a model at 300 K and deploying it across the range of 200 K to 400 K showed that the \ourmethod model is also significantly more transferable than the pairwise model.  This is intriguing because although the neural network is highly expressive, it is not obvious how it distills many-body contributions to the CG free energy that make the potential more transferable. This is even more surprising when considering the inherent noise in force-matching, which drastically reduces the model's ability to match its training data precisely. This stands in contrast to the construction of AA potentials, where forces and energies can be matched during training to extreme precision.

We furthermore applied the \ourmethod workflow to study supercritical methanol, benzene, and methane across a range of densities and temperatures to study transferability from a more broad perspective. These conditions span significant variation in structural ordering in the fluids. We found that while overall model accuracy is superior using single-state models, models transferred across thermodynamic states do not produce extreme levels of error, and in some instances appear to paradoxically produce somewhat lower error than single-state models. We also tested building force-matched models trained across multiple state points. Although such a workflow violates the thermodynamic assumptions of coarse-graining (as the CG free energy is a function of temperature and density), there was no difficulty in producing these multi-state models, and they performed very well. However, given the relatively surprising level of transferability of single-state \ourmethod models, the comparative advantage of training to multiple state points was not strong. Multi-state training did improve on the variance of the RDF error for higher density methanol and benzene.

\ADD{Finally, we estimated the speedup factor in of the \ourmethod models here over all-atom HIP-NN-TS potentials to assess the computational leverage of coarse-graining. This analysis leads to an estimate of 30x-150x speedup, with the large variance being due to the time step used in the CG model, which could in many cases have been pushed up to 5 fs. This shows that \ourmethod models do in fact accomplish key cost-reduction associated with coarse-graining. Simulation throughput can be drastically improved while retaining many characteristics of the underlying fine-scale (AA) system.}

\ADD{Although the \ourmethod models are able to reproduce the structure of the AA trajectories with high accuracy, they do not replicate the dynamics correctly. This is a known and widely-observed problem with coarse-grained models which occurs due to the smoothing of the potential energy surface and the removal of friction effects. Techniques to address this issue are currently in development.\cite{jin2023understanding, izvekov2006modeling, hijon2010mori, fritz2011multiscale, kinjo2007equation, lei2010direct, fao2011semi, davtyan2015dynamic, davtyan2016dynamic, izvekov2017microscopic, jung2017iterative, han2021constructing, izvekov2017mori}}

\STRIKE{One}\ADD{Another} area for future work is to explore transferability for cases of larger differences in structure, such as through phase transitions, e.g. between crystal and liquid. It might be that in this case, a multi-state training procedure shows stronger advantages. However, it is more difficult to automate the training data generation across thermodynamic phase changes, and this challenge would need to be addressed. Additionally, there are a wide variety of possible targets for coarse-graining, such as proteins, macromolecules, alloys, and liquid mixtures, which might be explored. Another future work possibility is to explore the potential for non-equilibrium coarse-grained simulations. Given the surprising transferability of these potentials, it may be possible to accurately model near-equilibrium conditions where temperatures and/or pressures evolve, either as a function of time or over space given by some boundary conditions. An exciting possibility is to incorporate additional thermodynamics into the free energy function, essentially applying the concept of thermodynamically consistent learning\cite{Rosenberger2022} to coarse-graining; such a concept has recently be introduced and explored for the coarse-graining of hexane\cite{duschatko2024}. With a wide range of recent improvements and ideas, machine learning based coarse-graining is poised to enable accurate simulations on large length and time scales across a wider ranges of thermodynamic conditions.

%% file: main_sections/data.tex
\section{Data Availability Statement}
The data that support the findings of this study are available from the corresponding authors upon reasonable request
and institutional approval.

%% file: si_sections/hyperparameter.tex
\section{\ADD{Neural network hyperparameter investigation}}

\ADD{In order to select hyperparameters for the \ourmethod models, a brief trial and error procedure was employed, beginning with values which had show success in previous HIP-NN-TS networks used for atomistic simulations. The final parameters were decided upon by balancing computational cost (both of training and evaluating the networks) with the networks' performance at matching forces in the loss function. The hyperparameters were fixed during an early phase of the study before any of the final data used in the study was generated.}
            
\ADD{However, we have additionally performed a post-hoc study using the following variants of the architecture:
\begin{itemize}
    \itemsep-0.5em
    \item Two interactions layers instead of 1
    \item More/fewer atom layers (5 or 2 instead of 3)
    \item More/fewer features (64 or 256 instead of 128)
    \item Stronger/weaker repulsive potential by a factor of 5x and a factor of 100x
    \item More/fewer sensitivity functions (10 or 30 instead of 20)
    \item MAE/RMSE loss only instead of combination loss
\end{itemize}
}

\ADD{
We performed these over four density/temperature combinations for benzene. With the exception of the strengthened repulsive potential, these alterations had little impact on the model's force-fitting capabilities or the resulting RDFs and ADFs. Except for the models with stronger repulsive potential, each RDF produced had TAE error below 0.81~\AA. For comparison, the largest RDF TAE observed for a benzene model in the main text is 0.80~\AA. That RDF is pictured in Figure \ref{figure: addl rdfs}(c). Overall, we did not observe significant correlation between a model's force RMSE and MAE scores and its performance at replicating the AA RDFs. 
}

\ADD{
In one of the four density/temperature combinations, the model with 5x increased strength for the repulsive potential resulted in an increase by about 30\% in force RMSE compared with the control. However, the resulting RDF TAE was just $0.34~\text{\AA}$ and the RDF TAEs for the other three temperature/density combinations were all under 0.6~\AA. Of the four models with 100x increased strength for the repulsive potential, in two cases (those for lower temperatures), the models were unable to produce stable MD. In the other two cases, though, the RDF TAEs were under 0.75~\AA. 
}

\ADD{
Based on these findings, we believe that there is a good deal of flexibility in selecting hyperparameters for the \ourmethod models. If any issues are encountered with exploding MD simulations, one should try decreasing the strength of the repulsive potential. The most important hyperparameter to set in HIP-NN is the initial distance parameters for the sensitivity functions. These must capture the range of interactions of interest in the data. Here, this was set to a conservative $2~\text{\AA}$ cutoff for all models. 
}

%% file: si_sections/repulsive_potential.tex
\section{Repulsive potential parameterization details} \label{appendix subsection: repulsive potential}

To help the HIP-NN-TS models handle pairs of beads whose distance is less than that of any pairs in the set of training data, a repulsive pair potential is added. The potential is of the form 
\begin{equation}
    E_\text{rep}(r) = E_0 e^{-ar}
\end{equation}
where $r$ is the pair distance, $E_0, a > 0$ are parameters chosen for each network, and $E_\text{rep}(r)$ is the contribution to the energy potential from the pair. The following values are used to calculate $E_0$ and $a$:
    
\begin{center}
    \begin{tabularx}{0.95\textwidth} { 
  | >{\raggedright\arraybackslash}X 
  | >{\raggedright\arraybackslash}X 
  | >{\raggedright\arraybackslash}X | }
        \hline
        \textbf{Variable name} &\textbf{Variable interpretation} & \textbf{Value(s) used} \\
        \hline
         $t$ & pair distance at which repulsive potential should be \STRIKE{minimal}\ADD{negligible} & \STRIKE{lowest pair distance in training data (outliers excluded)}\ADD{the center value of the first bin of the training data RDF for which the RDF value is greater than 0.01} \\
         \hline
         $s$ & strength of force on particle due to repulsive potential at $t-d$ & the mean force magnitude from the training data \\
         \hline
         $d$ & see preceding row & 0.15 (Angstroms) \\
         \hline
         $p$ & percentage of $s$ that remains at $t$ & 0.05 \\ 
        \hline
    \end{tabularx}
\end{center}

\noindent Based on these values, the values of $a$ and $E_0$ are calculated as
\begin{equation}
    a = \frac{1}{d} \ln\left(\frac{1}{p}\right) \text{  and  } E_0 = \frac{pse^{at}}{a}.
\end{equation}
Then the force on one particle from another particle at distance $r$ due to the repulsive potential is 
\begin{equation}
    F_{rep}(r) = -\frac{\partial E_{rep}}{\partial dr} (r) = aE_0e^{-ar},
\end{equation}
and one can verify that $F_{rep}(t) = ps$ and $F_{rep}(t-d) = s$. 

\begin{figure}[htbp]
        \includegraphics[align=c]{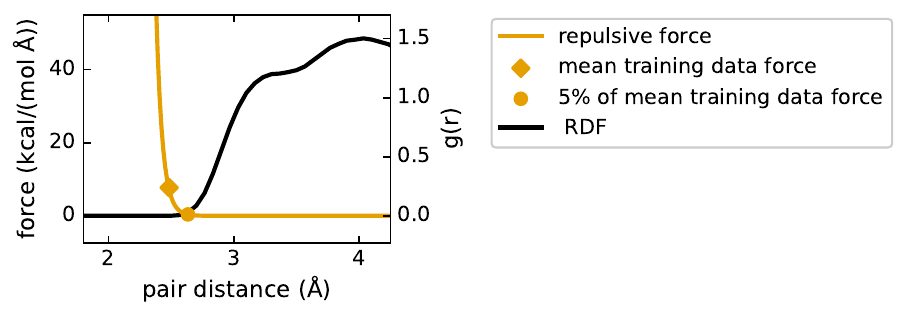}
        \caption{\ADD{Pairwise force generated by the repulsive potential for a single-state methanol model at 700 K, 1.03 g/cm$^3$ drawn over the RDF of the training data used for the model. }}   
        \label{figure: repulsive potential}
\end{figure}

\ADD{Figure~\ref{figure: repulsive potential} shows an example of the force generated by the repulsive potential used for the single-state methanol model at 700 K, 1.03 g/cm$^3$ on particle pairs at different distances. The RDF generated by the training data for this model, which demonstrates the prevalence of pairs at different distances is also included in the figure. The distance at which the repulsive potential generates force as strong as the average force from the training data is marked, as well as the point when the RDF values rise above 0.01 and the repulsive potential generates a force which is 5\% the average training data force. }

\clearpage

%% file: si_sections/training_statistics.tex
\section{Training statistics} \label{appendix subsection: training statistics}


Here the statistics of the trained HIP-NN-TS models are reported. Recall that the Total Loss is the sum of Force MAE and Force RMSE. The Force $R^2$ is also reported. Each statistic is displayed as the mean plus or minus the standard deviation across the five variants of that model which were trained.
\vspace{0.2in}

\begingroup
\let\clearpage\relax 
\include{si_sections/training_statistics_tables/methanol}
\include{si_sections/training_statistics_tables/benzene}
\include{si_sections/training_statistics_tables/methane}
\endgroup

\noindent Below are the statistics for the models trained for the comparison with the OpenMSCG method.
\vspace{0.2in}

\begingroup
\let\clearpage\relax
\include{si_sections/training_statistics_tables/methanol_mscg}
\endgroup

%% file: si_sections/training_statistics_tables/methanol.tex
\noindent
\textbf{\large Methanol}
\vspace{0.1in}

\noindent
\begin{footnotesize}
\begin{tabular}{lllllll}
\toprule
\toprule
 &  &  & Force MAE & Force RMSE & Total Loss & Force $R^2$ \\
Temp. & Density & Dataset &  &  &  &  \\
\midrule
\multirow[t]{9}{*}{600 K} & \multirow[t]{3}{*}{1.27 g/cm$^3$} & validation & $8.982\pm 0.002$ & $11.845\pm 0.002$ & $20.826\pm 0.003$ & $0.138\pm 0.000$ \\
 &  & testing & $9.010\pm 0.002$ & $11.862\pm 0.002$ & $20.872\pm 0.003$ & $0.140\pm 0.000$ \\
 &  & training & $8.963\pm 0.005$ & $11.819\pm 0.006$ & $20.782\pm 0.012$ & $0.143\pm 0.001$ \\
\cline{2-7}
 & \multirow[t]{3}{*}{1.03 g/cm$^3$} & validation & $6.560\pm 0.001$ & $8.877\pm 0.002$ & $15.436\pm 0.001$ & $0.153\pm 0.000$ \\
 &  & testing & $6.545\pm 0.001$ & $8.860\pm 0.001$ & $15.405\pm 0.001$ & $0.150\pm 0.000$ \\
 &  & training & $6.532\pm 0.003$ & $8.834\pm 0.002$ & $15.366\pm 0.004$ & $0.155\pm 0.000$ \\
\cline{2-7}
 & \multirow[t]{3}{*}{0.85 g/cm$^3$} & validation & $5.246\pm 0.001$ & $7.255\pm 0.002$ & $12.501\pm 0.003$ & $0.156\pm 0.001$ \\
 &  & testing & $5.246\pm 0.001$ & $7.260\pm 0.003$ & $12.506\pm 0.004$ & $0.161\pm 0.001$ \\
 &  & training & $5.263\pm 0.003$ & $7.288\pm 0.005$ & $12.551\pm 0.007$ & $0.161\pm 0.001$ \\
\cline{1-7}
\multirow[t]{9}{*}{700 K} & \multirow[t]{3}{*}{1.27 g/cm$^3$} & validation & $9.874\pm 0.001$ & $13.096\pm 0.001$ & $22.970\pm 0.002$ & $0.138\pm 0.000$ \\
 &  & testing & $9.867\pm 0.001$ & $13.084\pm 0.001$ & $22.951\pm 0.003$ & $0.135\pm 0.000$ \\
 &  & training & $9.872\pm 0.003$ & $13.079\pm 0.003$ & $22.950\pm 0.007$ & $0.137\pm 0.000$ \\
\cline{2-7}
 & \multirow[t]{3}{*}{1.03 g/cm$^3$} & validation & $7.146\pm 0.002$ & $9.742\pm 0.001$ & $16.888\pm 0.001$ & $0.147\pm 0.000$ \\
 &  & testing & $7.138\pm 0.003$ & $9.769\pm 0.001$ & $16.907\pm 0.002$ & $0.146\pm 0.000$ \\
 &  & training & $7.131\pm 0.004$ & $9.741\pm 0.002$ & $16.872\pm 0.006$ & $0.149\pm 0.000$ \\
\cline{2-7}
 & \multirow[t]{3}{*}{0.85 g/cm$^3$} & validation & $5.615\pm 0.001$ & $7.872\pm 0.001$ & $13.487\pm 0.002$ & $0.152\pm 0.000$ \\
 &  & testing & $5.640\pm 0.001$ & $7.912\pm 0.001$ & $13.552\pm 0.001$ & $0.150\pm 0.000$ \\
 &  & training & $5.624\pm 0.001$ & $7.906\pm 0.001$ & $13.529\pm 0.001$ & $0.151\pm 0.000$ \\
\cline{1-7}
\end{tabular}
\end{footnotesize}
\pagebreak

\noindent
\textbf{\large Methanol continued}
\vspace{0.1in}

\noindent
\begin{footnotesize}
\begin{tabular}{lllllll}
\midrule
 &  &  & Force MAE & Force RMSE & Total Loss & Force $R^2$ \\
Temp. & Density & Dataset &  &  &  &  \\
\midrule
\multirow[t]{9}{*}{800 K} & \multirow[t]{3}{*}{1.27 g/cm$^3$} & validation & $10.803\pm 0.002$ & $14.375\pm 0.003$ & $25.178\pm 0.004$ & $0.132\pm 0.000$ \\
 &  & testing & $10.704\pm 0.001$ & $14.272\pm 0.001$ & $24.976\pm 0.001$ & $0.136\pm 0.000$ \\
 &  & training & $10.762\pm 0.004$ & $14.330\pm 0.006$ & $25.092\pm 0.011$ & $0.137\pm 0.001$ \\
\cline{2-7}
 & \multirow[t]{3}{*}{1.03 g/cm$^3$} & validation & $7.686\pm 0.002$ & $10.557\pm 0.002$ & $18.243\pm 0.005$ & $0.146\pm 0.000$ \\
 &  & testing & $7.678\pm 0.002$ & $10.574\pm 0.002$ & $18.253\pm 0.003$ & $0.145\pm 0.000$ \\
 &  & training & $7.682\pm 0.003$ & $10.580\pm 0.003$ & $18.261\pm 0.006$ & $0.148\pm 0.001$ \\
\cline{2-7}
 & \multirow[t]{3}{*}{0.85 g/cm$^3$} & validation & $5.985\pm 0.002$ & $8.512\pm 0.002$ & $14.498\pm 0.002$ & $0.149\pm 0.000$ \\
 &  & testing & $5.979\pm 0.003$ & $8.481\pm 0.002$ & $14.461\pm 0.001$ & $0.148\pm 0.000$ \\
 &  & training & $5.979\pm 0.002$ & $8.499\pm 0.004$ & $14.478\pm 0.003$ & $0.151\pm 0.001$ \\
\cline{1-7}
\multirow[t]{3}{*}{Multi-state} & \multirow[t]{3}{*}{} & validation & $7.370\pm 0.001$ & $10.252\pm 0.002$ & $17.622\pm 0.001$ & $0.142\pm 0.000$ \\
 &  & testing & $7.849\pm 0.001$ & $10.910\pm 0.003$ & $18.760\pm 0.004$ & $0.143\pm 0.000$ \\
 &  & training & $7.539\pm 0.003$ & $10.485\pm 0.003$ & $18.024\pm 0.006$ & $0.141\pm 0.001$ \\
\cline{1-7}
\bottomrule
\end{tabular}
\end{footnotesize}

\pagebreak

%% file: si_sections/training_statistics_tables/benzene.tex
\noindent
\textbf{\large Benzene}
\vspace{0.1in}

\noindent
\begin{footnotesize}
\begin{tabular}{lllllll}
\toprule
\toprule
 &  &  & Force MAE & Force RMSE & Total Loss & Force $R^2$ \\
Temp. & Density & Dataset &  &  &  &  \\
\midrule
\multirow[t]{9}{*}{600 K} & \multirow[t]{3}{*}{1.06 g/cm$^3$} & validation & $6.710\pm 0.003$ & $9.084\pm 0.003$ & $15.794\pm 0.006$ & $0.161\pm 0.001$ \\
 &  & testing & $6.726\pm 0.003$ & $9.092\pm 0.004$ & $15.818\pm 0.006$ & $0.160\pm 0.001$ \\
 &  & training & $6.701\pm 0.004$ & $9.058\pm 0.004$ & $15.759\pm 0.007$ & $0.162\pm 0.001$ \\
\cline{2-7}
 & \multirow[t]{3}{*}{0.80 g/cm$^3$} & validation & $3.990\pm 0.002$ & $5.833\pm 0.002$ & $9.824\pm 0.002$ & $0.149\pm 0.001$ \\
 &  & testing & $4.030\pm 0.002$ & $5.900\pm 0.002$ & $9.929\pm 0.002$ & $0.146\pm 0.001$ \\
 &  & training & $4.014\pm 0.002$ & $5.879\pm 0.002$ & $9.892\pm 0.003$ & $0.147\pm 0.001$ \\
\cline{2-7}
 & \multirow[t]{3}{*}{0.61 g/cm$^3$} & validation & $2.843\pm 0.002$ & $4.491\pm 0.002$ & $7.334\pm 0.001$ & $0.127\pm 0.001$ \\
 &  & testing & $2.844\pm 0.002$ & $4.481\pm 0.002$ & $7.325\pm 0.001$ & $0.129\pm 0.001$ \\
 &  & training & $2.868\pm 0.006$ & $8.019\pm 1.734$ & $10.886\pm 1.740$ & $-1.865\pm 1.038$ \\
\cline{1-7}
\multirow[t]{9}{*}{700 K} & \multirow[t]{3}{*}{1.06 g/cm$^3$} & validation & $7.555\pm 0.002$ & $10.257\pm 0.002$ & $17.812\pm 0.003$ & $0.159\pm 0.000$ \\
 &  & testing & $7.558\pm 0.002$ & $10.284\pm 0.002$ & $17.842\pm 0.003$ & $0.161\pm 0.000$ \\
 &  & training & $7.562\pm 0.002$ & $10.275\pm 0.002$ & $17.837\pm 0.003$ & $0.163\pm 0.000$ \\
\cline{2-7}
 & \multirow[t]{3}{*}{0.80 g/cm$^3$} & validation & $4.482\pm 0.002$ & $6.612\pm 0.001$ & $11.094\pm 0.002$ & $0.147\pm 0.000$ \\
 &  & testing & $4.504\pm 0.002$ & $6.647\pm 0.002$ & $11.151\pm 0.002$ & $0.146\pm 0.000$ \\
 &  & training & $4.495\pm 0.002$ & $6.636\pm 0.002$ & $11.131\pm 0.003$ & $0.147\pm 0.001$ \\
\cline{2-7}
 & \multirow[t]{3}{*}{0.61 g/cm$^3$} & validation & $3.168\pm 0.003$ & $5.061\pm 0.003$ & $8.229\pm 0.002$ & $0.133\pm 0.001$ \\
 &  & testing & $3.144\pm 0.004$ & $5.016\pm 0.003$ & $8.161\pm 0.002$ & $0.131\pm 0.001$ \\
 &  & training & $3.173\pm 0.003$ & $6.907\pm 0.342$ & $10.080\pm 0.344$ & $-0.621\pm 0.159$ \\
\cline{1-7}
\multirow[t]{9}{*}{800 K} & \multirow[t]{3}{*}{1.06 g/cm$^3$} & validation & $8.410\pm 0.002$ & $11.480\pm 0.003$ & $19.890\pm 0.001$ & $0.161\pm 0.000$ \\
 &  & testing & $8.383\pm 0.001$ & $11.409\pm 0.003$ & $19.791\pm 0.002$ & $0.162\pm 0.000$ \\
 &  & training & $8.416\pm 0.002$ & $11.477\pm 0.005$ & $19.892\pm 0.006$ & $0.163\pm 0.001$ \\
\cline{2-7}
 & \multirow[t]{3}{*}{0.80 g/cm$^3$} & validation & $4.988\pm 0.003$ & $7.382\pm 0.002$ & $12.369\pm 0.001$ & $0.150\pm 0.000$ \\
 &  & testing & $4.991\pm 0.002$ & $7.370\pm 0.003$ & $12.361\pm 0.001$ & $0.152\pm 0.001$ \\
 &  & training & $4.998\pm 0.002$ & $7.405\pm 0.002$ & $12.403\pm 0.002$ & $0.151\pm 0.000$ \\
\cline{2-7}
 & \multirow[t]{3}{*}{0.61 g/cm$^3$} & validation & $3.518\pm 0.011$ & $5.699\pm 0.015$ & $9.217\pm 0.017$ & $0.129\pm 0.005$ \\
 &  & testing & $3.502\pm 0.011$ & $5.631\pm 0.007$ & $9.133\pm 0.011$ & $0.137\pm 0.002$ \\
 &  & training & $3.504\pm 0.011$ & $5.659\pm 0.007$ & $9.164\pm 0.014$ & $0.134\pm 0.002$ \\
\cline{1-7}
\end{tabular}
\end{footnotesize}
\pagebreak

\noindent
\textbf{\large Benzene continued}
\vspace{0.1in}

\noindent
\begin{footnotesize}
\begin{tabular}{lllllll}
\midrule
 &  &  & Force MAE & Force RMSE & Total Loss & Force $R^2$ \\
Temp. & Density & Dataset &  &  &  &  \\
\midrule
\multirow[t]{3}{*}{Multi-state} & \multirow[t]{3}{*}{} & validation & $4.862\pm 0.003$ & $7.381\pm 0.007$ & $12.243\pm 0.008$ & $0.145\pm 0.002$ \\
 &  & testing & $5.464\pm 0.009$ & $8.519\pm 0.534$ & $13.983\pm 0.543$ & $0.081\pm 0.120$ \\
 &  & training & $5.078\pm 0.006$ & $7.775\pm 0.181$ & $12.852\pm 0.187$ & $0.125\pm 0.041$ \\
\cline{1-7}
\bottomrule
\end{tabular}
\end{footnotesize}

\pagebreak

%% file: si_sections/training_statistics_tables/methane.tex
\noindent
\textbf{\large Methane}
\vspace{0.1in}

\noindent
\begin{footnotesize}
\begin{tabular}{lllllll}
\toprule
\toprule
 &  &  & Force MAE & Force RMSE & Total Loss & Force $R^2$ \\
Temp. & Density & Dataset &  &  &  &  \\
\midrule
\multirow[t]{9}{*}{300 K} & \multirow[t]{3}{*}{0.30 g/cm$^3$} & validation & $0.771\pm 0.001$ & $1.410\pm 0.001$ & $2.180\pm 0.000$ & $0.528\pm 0.001$ \\
 &  & testing & $0.765\pm 0.001$ & $1.388\pm 0.001$ & $2.153\pm 0.000$ & $0.534\pm 0.001$ \\
 &  & training & $0.765\pm 0.001$ & $1.390\pm 0.001$ & $2.156\pm 0.000$ & $0.533\pm 0.001$ \\
\cline{2-7}
 & \multirow[t]{3}{*}{0.16 g/cm$^3$} & validation & $0.429\pm 0.001$ & $0.978\pm 0.001$ & $1.408\pm 0.000$ & $0.535\pm 0.001$ \\
 &  & testing & $0.429\pm 0.001$ & $0.970\pm 0.001$ & $1.399\pm 0.000$ & $0.541\pm 0.001$ \\
 &  & training & $0.432\pm 0.001$ & $0.986\pm 0.001$ & $1.418\pm 0.000$ & $0.538\pm 0.001$ \\
\cline{2-7}
 & \multirow[t]{3}{*}{0.10 g/cm$^3$} & validation & $0.285\pm 0.000$ & $0.772\pm 0.000$ & $1.056\pm 0.000$ & $0.551\pm 0.000$ \\
 &  & testing & $0.289\pm 0.000$ & $0.807\pm 0.001$ & $1.097\pm 0.000$ & $0.535\pm 0.001$ \\
 &  & training & $0.283\pm 0.000$ & $0.780\pm 0.001$ & $1.063\pm 0.000$ & $0.545\pm 0.001$ \\
\cline{1-7}
\multirow[t]{9}{*}{500 K} & \multirow[t]{3}{*}{0.30 g/cm$^3$} & validation & $1.107\pm 0.002$ & $2.092\pm 0.001$ & $3.199\pm 0.001$ & $0.520\pm 0.001$ \\
 &  & testing & $1.098\pm 0.001$ & $2.096\pm 0.001$ & $3.194\pm 0.001$ & $0.527\pm 0.000$ \\
 &  & training & $1.095\pm 0.002$ & $2.080\pm 0.001$ & $3.175\pm 0.001$ & $0.527\pm 0.000$ \\
\cline{2-7}
 & \multirow[t]{3}{*}{0.16 g/cm$^3$} & validation & $0.568\pm 0.002$ & $1.347\pm 0.002$ & $1.914\pm 0.000$ & $0.538\pm 0.001$ \\
 &  & testing & $0.570\pm 0.002$ & $1.397\pm 0.002$ & $1.967\pm 0.001$ & $0.516\pm 0.001$ \\
 &  & training & $0.566\pm 0.002$ & $1.377\pm 0.002$ & $1.943\pm 0.001$ & $0.527\pm 0.001$ \\
\cline{2-7}
 & \multirow[t]{3}{*}{0.10 g/cm$^3$} & validation & $0.356\pm 0.001$ & $1.070\pm 0.001$ & $1.426\pm 0.000$ & $0.522\pm 0.001$ \\
 &  & testing & $0.350\pm 0.001$ & $1.051\pm 0.001$ & $1.400\pm 0.001$ & $0.529\pm 0.001$ \\
 &  & training & $0.348\pm 0.001$ & $1.037\pm 0.001$ & $1.385\pm 0.001$ & $0.532\pm 0.001$ \\
\cline{1-7}
\multirow[t]{9}{*}{700 K} & \multirow[t]{3}{*}{0.30 g/cm$^3$} & validation & $1.439\pm 0.002$ & $2.827\pm 0.002$ & $4.266\pm 0.001$ & $0.516\pm 0.001$ \\
 &  & testing & $1.458\pm 0.001$ & $2.873\pm 0.002$ & $4.332\pm 0.001$ & $0.511\pm 0.001$ \\
 &  & training & $1.439\pm 0.002$ & $2.815\pm 0.002$ & $4.254\pm 0.001$ & $0.521\pm 0.001$ \\
\cline{2-7}
 & \multirow[t]{3}{*}{0.16 g/cm$^3$} & validation & $0.714\pm 0.001$ & $1.826\pm 0.001$ & $2.540\pm 0.001$ & $0.520\pm 0.001$ \\
 &  & testing & $0.705\pm 0.001$ & $1.808\pm 0.001$ & $2.513\pm 0.000$ & $0.522\pm 0.000$ \\
 &  & training & $0.718\pm 0.001$ & $1.830\pm 0.001$ & $2.548\pm 0.001$ & $0.519\pm 0.001$ \\
\cline{2-7}
 & \multirow[t]{3}{*}{0.10 g/cm$^3$} & validation & $0.427\pm 0.001$ & $1.362\pm 0.001$ & $1.789\pm 0.000$ & $0.520\pm 0.001$ \\
 &  & testing & $0.429\pm 0.001$ & $1.356\pm 0.001$ & $1.785\pm 0.000$ & $0.519\pm 0.001$ \\
 &  & training & $0.429\pm 0.001$ & $1.360\pm 0.001$ & $1.790\pm 0.001$ & $0.527\pm 0.001$ \\
\cline{1-7}
\end{tabular}
\end{footnotesize}
\pagebreak

\noindent
\textbf{\large Methane continued}
\vspace{0.1in}

\noindent
\begin{footnotesize}
\begin{tabular}{lllllll}
\midrule
 &  &  & Force MAE & Force RMSE & Total Loss & Force $R^2$ \\
Temp. & Density & Dataset &  &  &  &  \\
\midrule
\multirow[t]{3}{*}{Multi-state} & \multirow[t]{3}{*}{} & validation & $0.632\pm 0.001$ & $1.539\pm 0.001$ & $2.172\pm 0.001$ & $0.523\pm 0.001$ \\
 &  & testing & $0.761\pm 0.001$ & $1.832\pm 0.002$ & $2.593\pm 0.001$ & $0.500\pm 0.001$ \\
 &  & training & $0.681\pm 0.001$ & $1.655\pm 0.001$ & $2.337\pm 0.001$ & $0.515\pm 0.001$ \\
\cline{1-7}
\bottomrule
\end{tabular}
\end{footnotesize}

\pagebreak

%% file: si_sections/training_statistics_tables/methanol_mscg.tex
\noindent
\textbf{\large Methanol}
\vspace{0.1in}

\noindent
\begin{footnotesize}
\begin{tabular}{llllll}
\toprule
\toprule
 &  & Force MAE & Force RMSE & Force $R^2$ & Total Loss \\
Temp. & Dataset &  &  &  &  \\
\midrule
\multirow[t]{3}{*}{200 K} & training & 2.5441 & 3.3669 & 0.2763 & 5.9110 \\
 & validation & 2.5507 & 3.3764 & 0.2736 & 5.9271 \\
 & testing & 2.4940 & 3.3089 & 0.3042 & 5.8029 \\
\cline{1-6}
\multirow[t]{3}{*}{220 K} & training & 2.6335 & 3.4968 & 0.2720 & 6.1302 \\
 & validation & 2.6375 & 3.5062 & 0.2708 & 6.1437 \\
 & testing & 2.6449 & 3.5087 & 0.2614 & 6.1536 \\
\cline{1-6}
\multirow[t]{3}{*}{240 K} & training & 2.7362 & 3.6470 & 0.2695 & 6.3832 \\
 & validation & 2.7248 & 3.6257 & 0.2712 & 6.3505 \\
 & testing & 2.7404 & 3.6462 & 0.2455 & 6.3866 \\
\cline{1-6}
\multirow[t]{3}{*}{260 K} & training & 2.8349 & 3.7875 & 0.2614 & 6.6224 \\
 & validation & 2.8323 & 3.7857 & 0.2623 & 6.6181 \\
 & testing & 2.7616 & 3.6527 & 0.2906 & 6.4143 \\
\cline{1-6}
\multirow[t]{3}{*}{280 K} & training & 2.9456 & 3.9430 & 0.2529 & 6.8886 \\
 & validation & 2.9440 & 3.9416 & 0.2530 & 6.8855 \\
 & testing & 2.8975 & 3.9620 & 0.2844 & 6.8595 \\
\cline{1-6}
\multirow[t]{3}{*}{300 K} & training & 3.0370 & 4.0829 & 0.2438 & 7.1199 \\
 & validation & 3.0356 & 4.0772 & 0.2474 & 7.1128 \\
 & testing & 2.9605 & 3.9363 & 0.2622 & 6.8968 \\
\cline{1-6}
\multirow[t]{3}{*}{320 K} & training & 3.1356 & 4.2195 & 0.2367 & 7.3551 \\
 & validation & 3.1319 & 4.2098 & 0.2355 & 7.3417 \\
 & testing & 3.0640 & 4.0750 & 0.2544 & 7.1390 \\
\cline{1-6}
\multirow[t]{3}{*}{340 K} & training & 3.2318 & 4.3665 & 0.2278 & 7.5983 \\
 & validation & 3.2320 & 4.3685 & 0.2240 & 7.6005 \\
 & testing & 3.2447 & 4.3918 & 0.2164 & 7.6365 \\
\cline{1-6}
\end{tabular}
\end{footnotesize}
\pagebreak

\noindent
\textbf{\large Methanol continued}
\vspace{0.1in}

\noindent
\begin{footnotesize}
\begin{tabular}{lllllll}
\midrule
 &  & Force MAE & Force RMSE & Force $R^2$ & Total Loss \\
Temp. & Dataset &  &  &  &  \\
\midrule
\multirow[t]{3}{*}{360 K} & training & 3.3178 & 4.4953 & 0.2221 & 7.8131 \\
 & validation & 3.3192 & 4.4978 & 0.2234 & 7.8169 \\
 & testing & 3.4481 & 4.6422 & 0.2309 & 8.0903 \\
\cline{1-6}
\multirow[t]{3}{*}{380 K} & training & 3.4135 & 4.6409 & 0.2119 & 8.0544 \\
 & validation & 3.4127 & 4.6332 & 0.2156 & 8.0459 \\
 & testing & 3.3035 & 4.4568 & 0.2265 & 7.7603 \\
\cline{1-6}
\multirow[t]{3}{*}{400 K} & training & 3.5046 & 4.7743 & 0.2067 & 8.2789 \\
 & validation & 3.5152 & 4.7927 & 0.2073 & 8.3079 \\
 & testing & 3.5246 & 4.8109 & 0.2198 & 8.3355 \\
\cline{1-6}
\bottomrule
\end{tabular}
\end{footnotesize}

\pagebreak

%% file: si_sections/data_size.tex
\section{Effects of training set size}

\begin{figure}[htbp]
    \includegraphics[align=c]{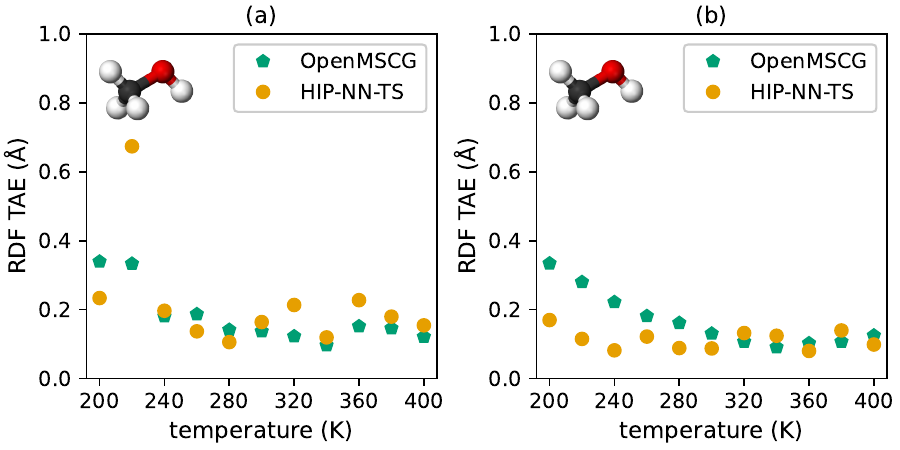}
    \caption{Results of 
    the methanol comparison study where both the \ourmethod and OpenMSCG models were trained using (a) 10 and (b) 100 frames of training data. 
    }
    \label{figure: training size openmscg}
\end{figure}

\begin{figure}[htbp]
    \includegraphics[align=c]{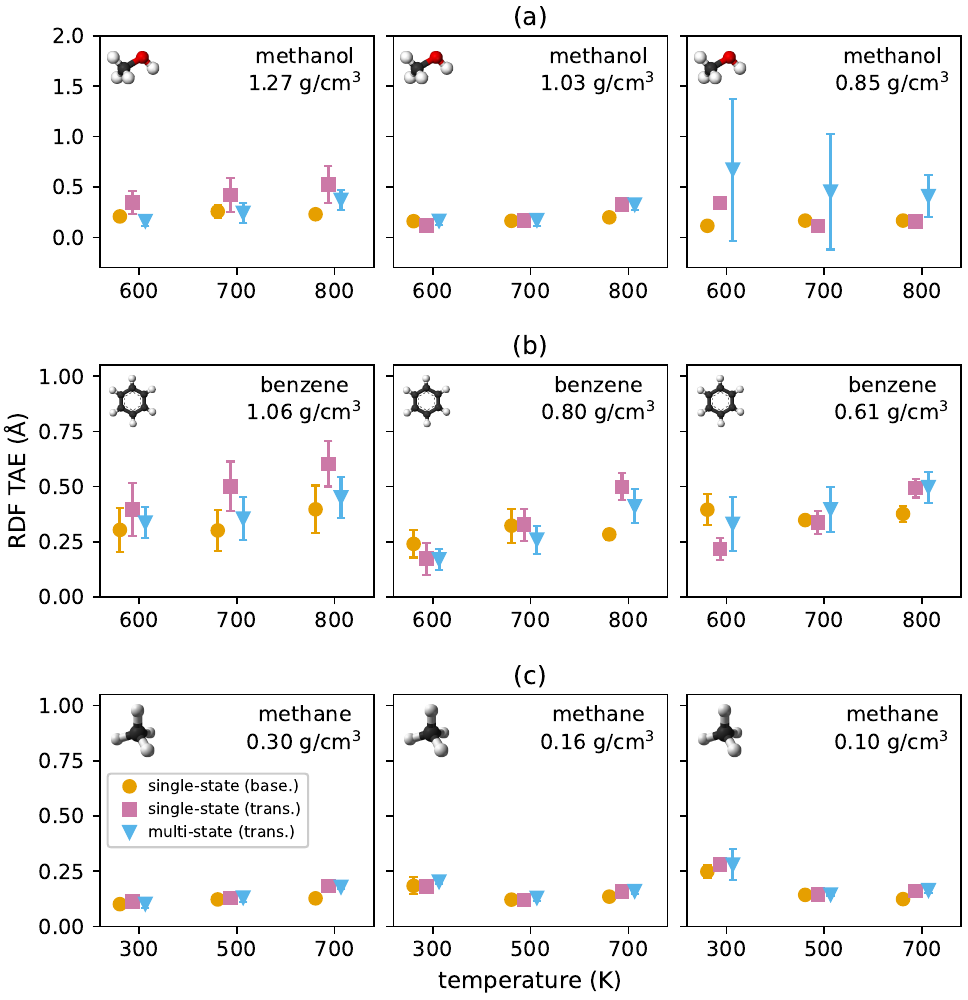}
    \caption{Results of the multi-molecule study when 100 frames rather than 1000 frames were used to train the various models for (a) methanol, (b) benzene, and (c) methane. Notice that the vertical scale on the methanol plot differs from that of the other two plots.}
    \label{figure: training size cross-molecular}
\end{figure}

We repeated the methods used in both the methanol comparison study and the multi-molecule study in the main text but with fewer frames in the \ourmethod and OpenMSCG model training data sets to investigate how the training data size affects the models' performance. The speed and automation of the developed workflows made training the 170+ additional \ourmethod models and running the 420+ additional MD runs needed for these analyses a feasible task. 

Figure~\ref{figure: training size openmscg} shows the results for the methanol comparison study. In this investigation, both the OpenMSCG models and the \ourmethod models were trained using only 10 (subfigure (a)) and 100 (subfigure (b)) frames in the training data set, as opposed to the 1,000 frames used in the results discussed in the main text. For 10 frames, both methods exhibit errors within two times the errors (i.e.~TAEs) of the original trial using 1,000 frames, with the exception of the \ourmethod model at 200 K, which has an error approximately four times the error for the 1,000 frames trial. Additionally, there is more fluctuation/noise in the results, indicating that the performance may be more sensitive to the random sample of training frames used. For the trial using 100 frames of training data, this fluctuation is noticeably decreased. In fact, the OpenMSCG results are only narrowly distinguishable from the results for 1,000 frames. Some fluctuation is still apparent in the results of the  \ourmethod models, although the errors are decreased in comparison with the 10 frame trial. These findings indicate that the performance of the OpenMSCG models may saturate earlier than the performance of the \ourmethod models, when provided access to more training data.

Figure~\ref{figure: training size cross-molecular} shows the results of the multi-molecule study when 100, rather than 1,000, frames of training data are used. In general, we observe higher error in some cases and a larger standard deviation (represented by the error bars) in many cases. For methanol (subfigure (a)), the errors are on average about a factor of two greater than the errors for the trial with 1,000 frames. The standard deviation of the errors is also 2-4 times greater in the highest density case, largely unchanged in the middle density case, and extremely large (up to 0.75 \AA) in the lowest density case. Closer investigation revealed that the cause of the high error in the lowest density case was an effect resembling a phase change in some of the MD runs. For benzene, the differences between the 100 frame trial and the 1,000 frame trial are less pronounced. Errors up to twice the errors of the 1,000 frame trial are observed in the 100 frame trial, but in most instances, the change is much less. Additionally, the standard deviations are up to twice as high in the 100 frame case, but again, in most cases the difference is not this great. Most notable is that the standard deviation for the single-state models is noticeably higher, particularly in the highest density case. Finally, the results for methane are shown in subfigure (c). The results in the main text for the methane models trained on 1,000 frames of data were remarkably consistent with extremely low error and standard deviation. The results show here for the trial with 100 frames are nearly identical, with the most noticeable difference being a higher standard deviation on the multi-state model for the low-temperature, low-density case. While the results for the methanol models were significantly improved by the addition of more training data, the methane models exhibit little change. This suggests that the needed number of frames of training data varies based on the geometry and/or size of the molecule.

%% file: si_sections/interpolation.tex
\section{Interpolation test}

\begin{figure}[htbp]
    \includegraphics[align=c]{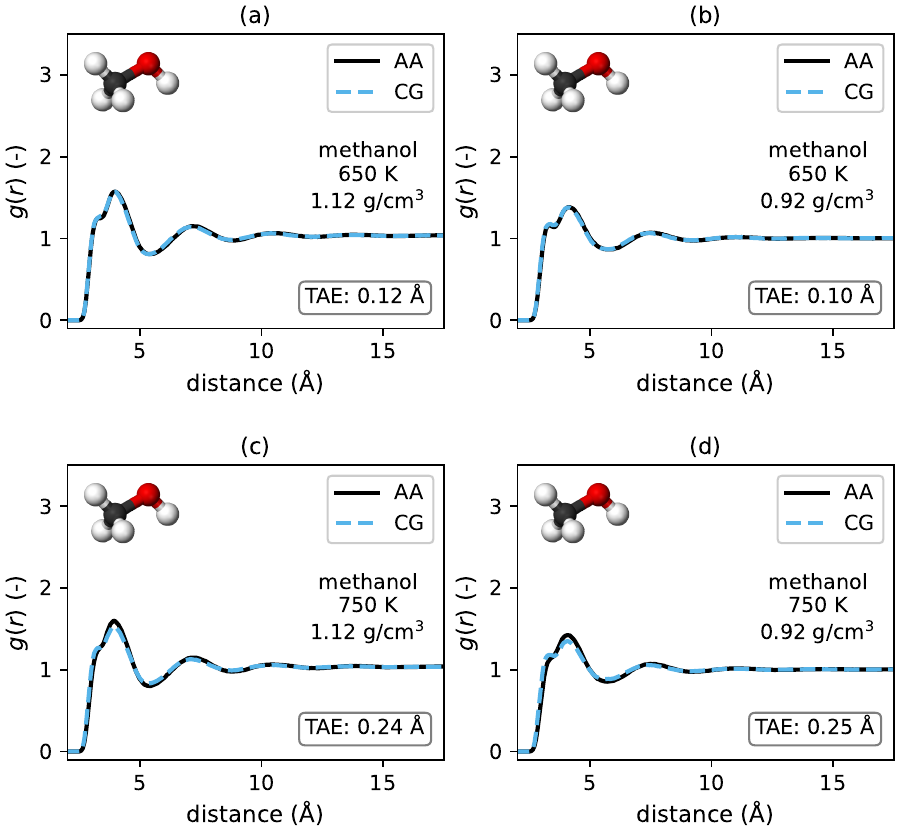}
    \caption{RDFs generated with AA MD compared with the multi-state \ourmethod models for methanol. The temperatures and densities at which these RDFs were generated were not included in the training data for the multi-state model.}
    \label{figure: interpolation test methanol}
\end{figure}

Figure \ref{figure: interpolation test methanol} shows the results of an interpolation test for the multi-state \ourmethod model trained for methanol in the cross-molecular study. This multi-state model was trained with data generated at nine state points representing each combination of three temperatures--600 K, 700 K, and 800 K--and three densities--1.27 g/cm$^3$, 1.03 g/cm$^3$, and 0.85 g/cm$^3$. The purpose of the test was to determine whether the model had simply learned how to perform at those nine state points or if it is able to extend its performance to intermediate state points (temperature between 600 K and 800 K and density between 1.27 g/cm$^3$ and 0.85 g/cm$^3$). 

The four state points chosen for this test were at each combination of the temperatures 650 K and 750 K and densities 1.12 g/cm$^3$ and 0.92 g/cm$^3$. For each test state point, AA MD was run in the same manner used to generate training data as described in the Training subsection of the main text. This AA data was not used for model training, however, but solely to generate AA RDFs. Randomly selected frames from the AA trajectories were also used as initial starting points for the CG MD. The RDFs generated from running CG MD using the methanol multi-state model are also shown in the figure. The resulting TAEs range from 0.10 \AA~to 0.25 \AA, perfectly in line with the results for the state points represented in the model training data which ranged from about 0.10 \AA~ to about 0.45~ \AA. The general trend of the methanol multi-state model performing slightly better at lower temperatures, which was observed in analysis of the cross-molecular study results in the main text, can also be seen clearly here. Based on these findings, we expect that the multi-state model would perform similarly well on any state points within the range of the model training data, with slightly better results at lower temperatures. 

%% file: si_sections/addl_rdfs.tex
\section{\ADD{Additional RDF and ADF plots}}

\begin{figure}[htbp]
    \includegraphics[align=c,scale=1]{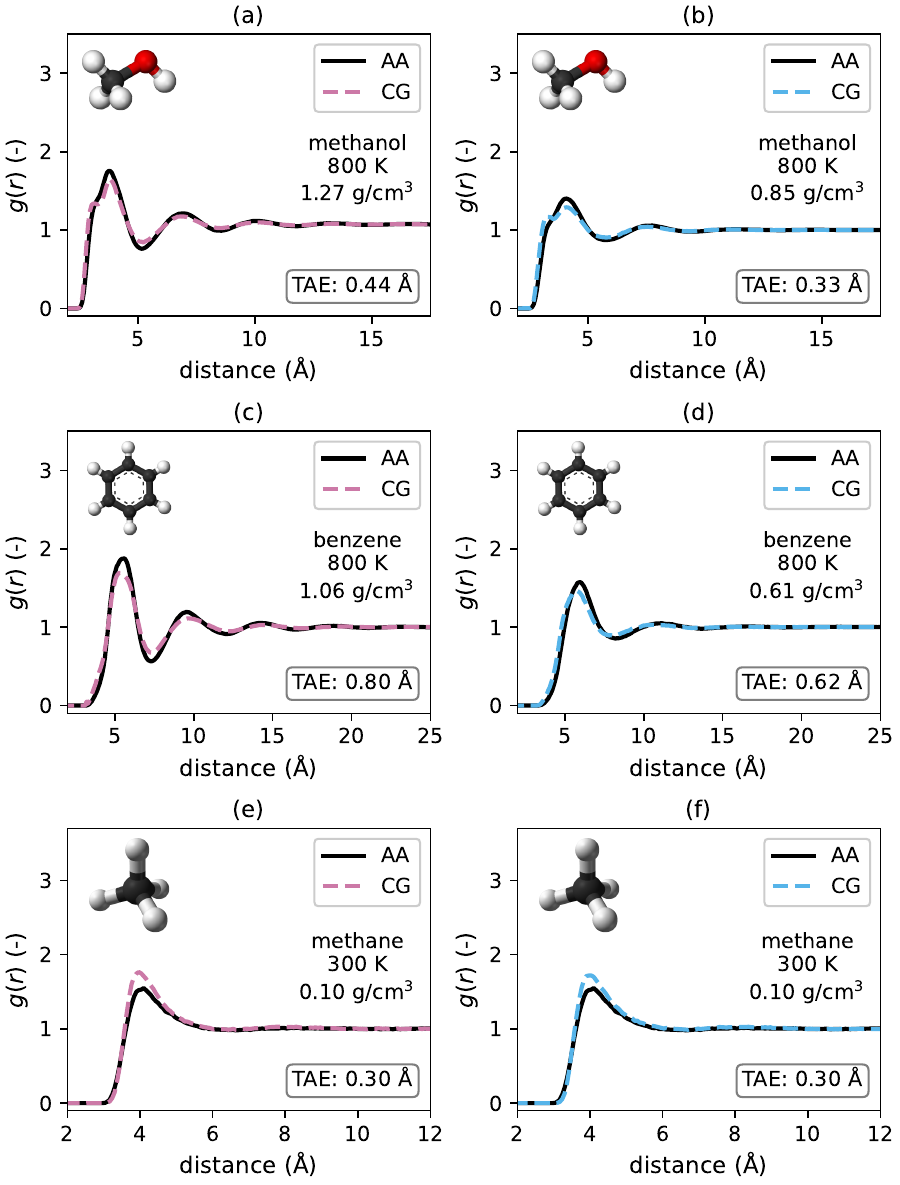}
    \caption{\ADD{Some of the RDFs with the highest TAE values among the 405 comparisons which make up Figure 7 in the main text. Subfigures (a), (c), and (e) show comparisons of single-state transferability test models and subfigures (b), (d), and (f) show RDFs from multi-state transferability test models.}}
    \label{figure: addl rdfs}
\end{figure}

\begin{figure}[htbp]
    \includegraphics[align=c]{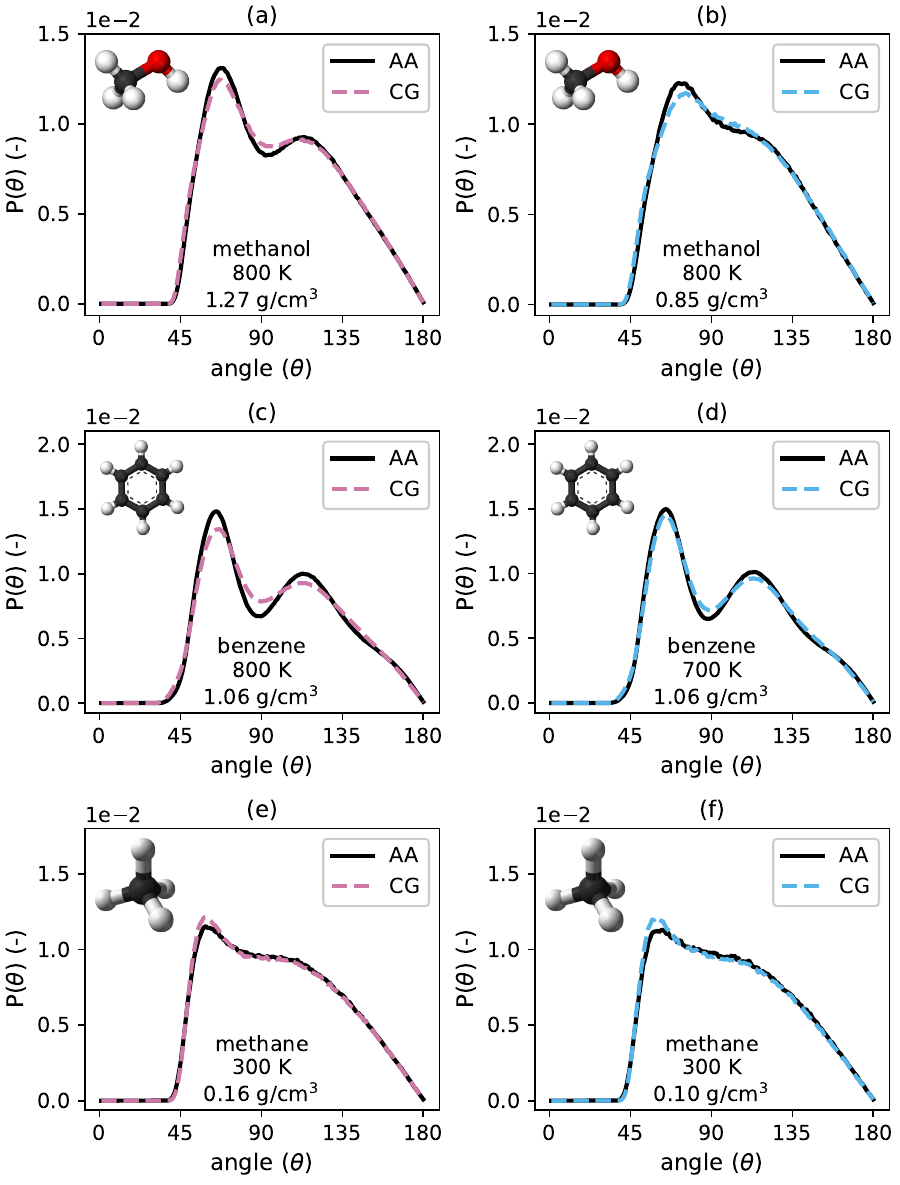}
    \caption{\ADD{Some of the ADFs with the highest MAE values among the 405 comparisons which make up Figure 10 in the main text. Subfigures (a), (c), and (e) show comparisons of single-state transferability test models and subfigures (b), (d), and (f) show RDFs from multi-state transferability test models.}}
    \label{figure: addl adfs}
\end{figure}

%% file: main.bbl
\providecommand{\latin}[1]{#1}
\makeatletter
\providecommand{\doi}
  {\begingroup\let\do\@makeother\dospecials
  \catcode`\{=1 \catcode`\}=2 \doi@aux}
\providecommand{\doi@aux}[1]{\endgroup\texttt{#1}}
\makeatother
\providecommand*\mcitethebibliography{\thebibliography}
\csname @ifundefined\endcsname{endmcitethebibliography}
  {\let\endmcitethebibliography\endthebibliography}{}
\begin{mcitethebibliography}{109}
\providecommand*\natexlab[1]{#1}
\providecommand*\mciteSetBstSublistMode[1]{}
\providecommand*\mciteSetBstMaxWidthForm[2]{}
\providecommand*\mciteBstWouldAddEndPuncttrue
  {\def\EndOfBibitem{\unskip.}}
\providecommand*\mciteBstWouldAddEndPunctfalse
  {\let\EndOfBibitem\relax}
\providecommand*\mciteSetBstMidEndSepPunct[3]{}
\providecommand*\mciteSetBstSublistLabelBeginEnd[3]{}
\providecommand*\EndOfBibitem{}
\mciteSetBstSublistMode{f}
\mciteSetBstMaxWidthForm{subitem}{(\alph{mcitesubitemcount})}
\mciteSetBstSublistLabelBeginEnd
  {\mcitemaxwidthsubitemform\space}
  {\relax}
  {\relax}

\bibitem[Izvekov and Voth(2005)Izvekov, and Voth]{Voth2005}
Izvekov,~S.; Voth,~G.~A. Multiscale coarse graining of liquid-state systems.
  \emph{J. Chem. Phys.} \textbf{2005}, \emph{123}, 134105\relax
\mciteBstWouldAddEndPuncttrue
\mciteSetBstMidEndSepPunct{\mcitedefaultmidpunct}
{\mcitedefaultendpunct}{\mcitedefaultseppunct}\relax
\EndOfBibitem
\bibitem[Lu \latin{et~al.}(2021)Lu, Wu, Ghoreishi, Chen, Wang, Damm, Ross,
  Dahlgren, Russell, Von~Bargen, Abel, Friesner, and Harder]{Lu2021OPLS}
Lu,~C.; Wu,~C.; Ghoreishi,~D.; Chen,~W.; Wang,~L.; Damm,~W.; Ross,~G.~A.;
  Dahlgren,~M.~K.; Russell,~E.; Von~Bargen,~C.~D.; Abel,~R.; Friesner,~R.~A.;
  Harder,~E.~D. OPLS4: Improving Force Field Accuracy on Challenging Regimes of
  Chemical Space. \emph{J. Chem. Theory Comput.} \textbf{2021}, \emph{17},
  4291--4300\relax
\mciteBstWouldAddEndPuncttrue
\mciteSetBstMidEndSepPunct{\mcitedefaultmidpunct}
{\mcitedefaultendpunct}{\mcitedefaultseppunct}\relax
\EndOfBibitem
\bibitem[Unke \latin{et~al.}(2021)Unke, Chmiela, Sauceda, Gastegger, Poltavsky,
  Schütt, Tkatchenko, and Müller]{Unke2021}
Unke,~O.~T.; Chmiela,~S.; Sauceda,~H.~E.; Gastegger,~M.; Poltavsky,~I.;
  Schütt,~K.~T.; Tkatchenko,~A.; Müller,~K.-R. Machine Learning Force Fields.
  \emph{Chem. Rev.} \textbf{2021}, \emph{121}, 10142--10186\relax
\mciteBstWouldAddEndPuncttrue
\mciteSetBstMidEndSepPunct{\mcitedefaultmidpunct}
{\mcitedefaultendpunct}{\mcitedefaultseppunct}\relax
\EndOfBibitem
\bibitem[Pol{\^e}to and Lemkul(2022)Pol{\^e}to, and Lemkul]{Poleto2022}
Pol{\^e}to,~M.~D.; Lemkul,~J.~A. Integration of experimental data and use of
  automated fitting methods in developing protein force fields. \emph{Comm.
  Chem.} \textbf{2022}, \emph{5}, 38\relax
\mciteBstWouldAddEndPuncttrue
\mciteSetBstMidEndSepPunct{\mcitedefaultmidpunct}
{\mcitedefaultendpunct}{\mcitedefaultseppunct}\relax
\EndOfBibitem
\bibitem[Zongo \latin{et~al.}(2022)Zongo, B{\'e}land, and
  Ouellet-Plamondon]{Zongo2022first}
Zongo,~K.; B{\'e}land,~L.; Ouellet-Plamondon,~C. First-principles database for
  fitting a machine-learning silicon interatomic force field. \emph{MRS
  Advances} \textbf{2022}, \emph{7}, 39--47\relax
\mciteBstWouldAddEndPuncttrue
\mciteSetBstMidEndSepPunct{\mcitedefaultmidpunct}
{\mcitedefaultendpunct}{\mcitedefaultseppunct}\relax
\EndOfBibitem
\bibitem[Matin \latin{et~al.}(2024)Matin, Allen, Smith, Lubbers, Jadrich,
  Messerly, Nebgen, Li, Tretiak, and Barros]{Matin2024}
Matin,~S.; Allen,~A. E.~A.; Smith,~J.; Lubbers,~N.; Jadrich,~R.~B.;
  Messerly,~R.; Nebgen,~B.; Li,~Y.~W.; Tretiak,~S.; Barros,~K. Machine Learning
  Potentials with the Iterative Boltzmann Inversion: Training to Experiment.
  \emph{J. Chem. Theory Comput.} \textbf{2024}, \emph{20}, 1274--1281\relax
\mciteBstWouldAddEndPuncttrue
\mciteSetBstMidEndSepPunct{\mcitedefaultmidpunct}
{\mcitedefaultendpunct}{\mcitedefaultseppunct}\relax
\EndOfBibitem
\bibitem[Botu \latin{et~al.}(2017)Botu, Batra, Chapman, and
  Ramprasad]{Ramprasad2017}
Botu,~V.; Batra,~R.; Chapman,~J.; Ramprasad,~R. Machine Learning Force Fields:
  Construction, Validation, and Outlook. \emph{J. Phys. Chem. C} \textbf{2017},
  \emph{121}, 511--522\relax
\mciteBstWouldAddEndPuncttrue
\mciteSetBstMidEndSepPunct{\mcitedefaultmidpunct}
{\mcitedefaultendpunct}{\mcitedefaultseppunct}\relax
\EndOfBibitem
\bibitem[Wu \latin{et~al.}(2023)Wu, Yang, Zhao, Li, Lu, Xie, and Yan]{Wu2023}
Wu,~S.; Yang,~X.; Zhao,~X.; Li,~Z.; Lu,~M.; Xie,~X.; Yan,~J. Applications and
  Advances in Machine Learning Force Fields. \emph{J. Chem. Inf. Model.}
  \textbf{2023}, \emph{63}, 6972--6985\relax
\mciteBstWouldAddEndPuncttrue
\mciteSetBstMidEndSepPunct{\mcitedefaultmidpunct}
{\mcitedefaultendpunct}{\mcitedefaultseppunct}\relax
\EndOfBibitem
\bibitem[Rosenberger \latin{et~al.}(2021)Rosenberger, Smith, and
  Garcia]{Rosenberger2021}
Rosenberger,~D.; Smith,~J.~S.; Garcia,~A.~E. Modeling of Peptides with
  Classical and Novel Machine Learning Force Fields: A Comparison. \emph{J.
  Phys. Chem. B} \textbf{2021}, \emph{125}, 3598--3612\relax
\mciteBstWouldAddEndPuncttrue
\mciteSetBstMidEndSepPunct{\mcitedefaultmidpunct}
{\mcitedefaultendpunct}{\mcitedefaultseppunct}\relax
\EndOfBibitem
\bibitem[Frenkel and Smit(1996)Frenkel, and Smit]{FrenkelSmit96}
Frenkel,~D.; Smit,~B. \emph{Understanding molecular simulation: From algorithms
  to application}; Academic Press: NY, 1996\relax
\mciteBstWouldAddEndPuncttrue
\mciteSetBstMidEndSepPunct{\mcitedefaultmidpunct}
{\mcitedefaultendpunct}{\mcitedefaultseppunct}\relax
\EndOfBibitem
\bibitem[Fedik \latin{et~al.}(2022)Fedik, Zubatyuk, Kulichenko, Lubbers, Smith,
  Nebgen, Messerly, Li, Boldyrev, Barros, \latin{et~al.} others]{Fedik2022}
Fedik,~N.; Zubatyuk,~R.; Kulichenko,~M.; Lubbers,~N.; Smith,~J.~S.; Nebgen,~B.;
  Messerly,~R.; Li,~Y.~W.; Boldyrev,~A.~I.; Barros,~K.; others Extending
  machine learning beyond interatomic potentials for predicting molecular
  properties. \emph{Nat. Rev. Chem.} \textbf{2022}, \emph{6}, 653--672\relax
\mciteBstWouldAddEndPuncttrue
\mciteSetBstMidEndSepPunct{\mcitedefaultmidpunct}
{\mcitedefaultendpunct}{\mcitedefaultseppunct}\relax
\EndOfBibitem
\bibitem[Perez \latin{et~al.}(2009)Perez, Uberuaga, Shim, Amar, and
  Voter]{perez2009accelerated}
Perez,~D.; Uberuaga,~B.~P.; Shim,~Y.; Amar,~J.~G.; Voter,~A.~F. Accelerated
  molecular dynamics methods: introduction and recent developments.
  \emph{Annual Reports in computational chemistry} \textbf{2009}, \emph{5},
  79--98\relax
\mciteBstWouldAddEndPuncttrue
\mciteSetBstMidEndSepPunct{\mcitedefaultmidpunct}
{\mcitedefaultendpunct}{\mcitedefaultseppunct}\relax
\EndOfBibitem
\bibitem[H{\'e}nin \latin{et~al.}(2022)H{\'e}nin, Leli{\`e}vre, Shirts,
  Valsson, and Delemotte]{henin2022enhanced}
H{\'e}nin,~J.; Leli{\`e}vre,~T.; Shirts,~M.~R.; Valsson,~O.; Delemotte,~L.
  Enhanced Sampling Methods for Molecular Dynamics Simulations [Article v1.0].
  \emph{Living Journal of Computational Molecular Science} \textbf{2022},
  \emph{4}, 1583\relax
\mciteBstWouldAddEndPuncttrue
\mciteSetBstMidEndSepPunct{\mcitedefaultmidpunct}
{\mcitedefaultendpunct}{\mcitedefaultseppunct}\relax
\EndOfBibitem
\bibitem[Rudzinski and Noid(2014)Rudzinski, and Noid]{Rudzinski2014}
Rudzinski,~J.~F.; Noid,~W.~G. Investigation of Coarse-Grained Mappings via an
  Iterative Generalized Yvon–Born–Green Method. \emph{J. Phys. Chem. B}
  \textbf{2014}, \emph{118}, 8295--8312\relax
\mciteBstWouldAddEndPuncttrue
\mciteSetBstMidEndSepPunct{\mcitedefaultmidpunct}
{\mcitedefaultendpunct}{\mcitedefaultseppunct}\relax
\EndOfBibitem
\bibitem[Shih \latin{et~al.}(2006)Shih, Arkhipov, Freddolino, , and
  Schulten]{Schulten2006}
Shih,~A.~Y.; Arkhipov,~A.; Freddolino,~P.~L.; ; Schulten,~K. Coarse Grained
  Protein-Lipid Model with Application to Lipoprotein Particles. \emph{J. Phys.
  Chem. B} \textbf{2006}, \emph{110}, 3674--3684\relax
\mciteBstWouldAddEndPuncttrue
\mciteSetBstMidEndSepPunct{\mcitedefaultmidpunct}
{\mcitedefaultendpunct}{\mcitedefaultseppunct}\relax
\EndOfBibitem
\bibitem[Saunders and Voth(2013)Saunders, and Voth]{Voth2013}
Saunders,~M.~G.; Voth,~G.~A. Coarse-Graining Methods for Computational Biology.
  \emph{Annu. Rev. Biophys.} \textbf{2013}, \emph{42}, 73--93\relax
\mciteBstWouldAddEndPuncttrue
\mciteSetBstMidEndSepPunct{\mcitedefaultmidpunct}
{\mcitedefaultendpunct}{\mcitedefaultseppunct}\relax
\EndOfBibitem
\bibitem[Lu \latin{et~al.}(2013)Lu, Dama, and Voth]{VothandLu2013}
Lu,~L.; Dama,~J.~F.; Voth,~G.~A. Fitting coarse-grained distribution functions
  through an iterative force-matching method. \emph{J. Chem. Phys.}
  \textbf{2013}, \emph{139}, 121906\relax
\mciteBstWouldAddEndPuncttrue
\mciteSetBstMidEndSepPunct{\mcitedefaultmidpunct}
{\mcitedefaultendpunct}{\mcitedefaultseppunct}\relax
\EndOfBibitem
\bibitem[Noid(2013)]{Noid2013}
Noid,~W.~G. Perspective: {Coarse-grained} models for biomolecular systems.
  \emph{J. Chem. Phys.} \textbf{2013}, \emph{139}, 090901\relax
\mciteBstWouldAddEndPuncttrue
\mciteSetBstMidEndSepPunct{\mcitedefaultmidpunct}
{\mcitedefaultendpunct}{\mcitedefaultseppunct}\relax
\EndOfBibitem
\bibitem[Jin \latin{et~al.}(2022)Jin, Pak, Durumeric, Loose, and
  Voth]{Jin2022bottom}
Jin,~J.; Pak,~A.~J.; Durumeric,~A.~E.; Loose,~T.~D.; Voth,~G.~A. Bottom-up
  coarse-graining: Principles and perspectives. \emph{J. Chem. Theory Comput.}
  \textbf{2022}, \emph{18}, 5759--5791\relax
\mciteBstWouldAddEndPuncttrue
\mciteSetBstMidEndSepPunct{\mcitedefaultmidpunct}
{\mcitedefaultendpunct}{\mcitedefaultseppunct}\relax
\EndOfBibitem
\bibitem[Dama \latin{et~al.}(2013)Dama, Sinitskiy, McCullagh, Weare, Roux,
  Dinner, and Voth]{Dama2013}
Dama,~J.~F.; Sinitskiy,~A.~V.; McCullagh,~M.; Weare,~J.; Roux,~B.;
  Dinner,~A.~R.; Voth,~G.~A. The Theory of Ultra-Coarse-Graining. {1.}
  {General} Principles. \emph{J. Chem. Theory Comput.} \textbf{2013}, \emph{9},
  2466--2480\relax
\mciteBstWouldAddEndPuncttrue
\mciteSetBstMidEndSepPunct{\mcitedefaultmidpunct}
{\mcitedefaultendpunct}{\mcitedefaultseppunct}\relax
\EndOfBibitem
\bibitem[Davtyan \latin{et~al.}(2014)Davtyan, Dama, Sinitskiy, and
  Voth]{VothUCG2014}
Davtyan,~A.; Dama,~J.~F.; Sinitskiy,~A.~V.; Voth,~G.~A. The Theory of
  Ultra-Coarse-Graining. {2.} {Numerical} Implementation. \emph{J. Chem. Theory
  Comput.} \textbf{2014}, \emph{10}, 5265--5275\relax
\mciteBstWouldAddEndPuncttrue
\mciteSetBstMidEndSepPunct{\mcitedefaultmidpunct}
{\mcitedefaultendpunct}{\mcitedefaultseppunct}\relax
\EndOfBibitem
\bibitem[Brennan \latin{et~al.}(2014)Brennan, Lísal, Moore, Izvekov,
  Schweigert, and Larentzos]{Brennan2014}
Brennan,~J.~K.; Lísal,~M.; Moore,~J.~D.; Izvekov,~S.; Schweigert,~I.~V.;
  Larentzos,~J.~P. Coarse-Grain Model Simulations of Nonequilibrium Dynamics in
  Heterogeneous Materials. \emph{J. Phys. Chem. Lett.} \textbf{2014}, \emph{5},
  2144--2149\relax
\mciteBstWouldAddEndPuncttrue
\mciteSetBstMidEndSepPunct{\mcitedefaultmidpunct}
{\mcitedefaultendpunct}{\mcitedefaultseppunct}\relax
\EndOfBibitem
\bibitem[Schilling(2022)]{Schilling2022noneqCG}
Schilling,~T. Coarse-grained modelling out of equilibrium. \emph{Phys. Rep.}
  \textbf{2022}, \emph{972}, 1--45\relax
\mciteBstWouldAddEndPuncttrue
\mciteSetBstMidEndSepPunct{\mcitedefaultmidpunct}
{\mcitedefaultendpunct}{\mcitedefaultseppunct}\relax
\EndOfBibitem
\bibitem[Marrink \latin{et~al.}(2007)Marrink, Risselada, Yefimov, Tieleman, and
  de~Vries]{cg-marrink}
Marrink,~S.~J.; Risselada,~H.~J.; Yefimov,~S.; Tieleman,~D.~P.; de~Vries,~A.~H.
  The MARTINI Force Field: Coarse Grained Model for Biomolecular Simulations.
  \emph{J. Phys. Chem. B} \textbf{2007}, \emph{111}, 7812--7824, PMID:
  17569554\relax
\mciteBstWouldAddEndPuncttrue
\mciteSetBstMidEndSepPunct{\mcitedefaultmidpunct}
{\mcitedefaultendpunct}{\mcitedefaultseppunct}\relax
\EndOfBibitem
\bibitem[Noid(2023)]{Noid2023}
Noid,~W.~G. Perspective: Advances, Challenges, and Insight for Predictive
  Coarse-Grained Models. \emph{J. Phys. Chem. B} \textbf{2023}, \emph{127},
  4174--4207\relax
\mciteBstWouldAddEndPuncttrue
\mciteSetBstMidEndSepPunct{\mcitedefaultmidpunct}
{\mcitedefaultendpunct}{\mcitedefaultseppunct}\relax
\EndOfBibitem
\bibitem[Palma~Banos \latin{et~al.}(2024)Palma~Banos, Popov, and
  Hernandez]{Palma2024}
Palma~Banos,~M.; Popov,~A.~V.; Hernandez,~R. Representability and Dynamical
  Consistency in Coarse-Grained Models. \emph{J. Phys. Chem. B} \textbf{2024},
  \emph{128}, 1506--1514\relax
\mciteBstWouldAddEndPuncttrue
\mciteSetBstMidEndSepPunct{\mcitedefaultmidpunct}
{\mcitedefaultendpunct}{\mcitedefaultseppunct}\relax
\EndOfBibitem
\bibitem[Craven \latin{et~al.}(2014)Craven, Popov, and Hernandez]{craven14b}
Craven,~G.~T.; Popov,~A.~V.; Hernandez,~R. Structure of a tractable stochastic
  mimic of soft particles. \emph{Soft Matter} \textbf{2014}, \emph{10},
  5350--5361\relax
\mciteBstWouldAddEndPuncttrue
\mciteSetBstMidEndSepPunct{\mcitedefaultmidpunct}
{\mcitedefaultendpunct}{\mcitedefaultseppunct}\relax
\EndOfBibitem
\bibitem[Moore \latin{et~al.}(2014)Moore, Iacovella, and McCabe]{Moore2014}
Moore,~T.~C.; Iacovella,~C.~R.; McCabe,~C. Derivation of coarse-grained
  potentials via multistate iterative {B}oltzmann inversion. \emph{J. Chem.
  Phys.} \textbf{2014}, \emph{140}, 224104\relax
\mciteBstWouldAddEndPuncttrue
\mciteSetBstMidEndSepPunct{\mcitedefaultmidpunct}
{\mcitedefaultendpunct}{\mcitedefaultseppunct}\relax
\EndOfBibitem
\bibitem[Dunn and Noid(2015)Dunn, and Noid]{Noid2015}
Dunn,~N. J.~H.; Noid,~W.~G. {Bottom-up coarse-grained models that accurately
  describe the structure, pressure, and compressibility of molecular liquids}.
  \emph{J. Chem. Phys.} \textbf{2015}, \emph{143}, 243148\relax
\mciteBstWouldAddEndPuncttrue
\mciteSetBstMidEndSepPunct{\mcitedefaultmidpunct}
{\mcitedefaultendpunct}{\mcitedefaultseppunct}\relax
\EndOfBibitem
\bibitem[Moradzadeh and Aluru(2019)Moradzadeh, and Aluru]{Moradzadeh2019}
Moradzadeh,~A.; Aluru,~N.~R. Transfer-Learning-Based Coarse-Graining Method for
  Simple Fluids: Toward Deep Inverse Liquid-State Theory. \emph{J. Phys. Chem.
  Lett.} \textbf{2019}, \emph{10}, 1242--1250\relax
\mciteBstWouldAddEndPuncttrue
\mciteSetBstMidEndSepPunct{\mcitedefaultmidpunct}
{\mcitedefaultendpunct}{\mcitedefaultseppunct}\relax
\EndOfBibitem
\bibitem[Narros \latin{et~al.}(2010)Narros, Moreno, and Likos]{Likos2010}
Narros,~A.; Moreno,~A.~J.; Likos,~C.~N. Influence of topology on effective
  potentials: coarse-graining ring polymers. \emph{Soft Matter} \textbf{2010},
  \emph{6}, 2435--2441\relax
\mciteBstWouldAddEndPuncttrue
\mciteSetBstMidEndSepPunct{\mcitedefaultmidpunct}
{\mcitedefaultendpunct}{\mcitedefaultseppunct}\relax
\EndOfBibitem
\bibitem[Craven \latin{et~al.}(2014)Craven, Popov, and Hernandez]{craven14d}
Craven,~G.~T.; Popov,~A.~V.; Hernandez,~R. Effective Surface Coverage of
  Coarse-Grained Soft Matter. \emph{J. Phys. Chem. B} \textbf{2014},
  \emph{118}, 14092--14102\relax
\mciteBstWouldAddEndPuncttrue
\mciteSetBstMidEndSepPunct{\mcitedefaultmidpunct}
{\mcitedefaultendpunct}{\mcitedefaultseppunct}\relax
\EndOfBibitem
\bibitem[Zhang and Wang(2013)Zhang, and Wang]{Wang2013}
Zhang,~P.; Wang,~Q. Solvent entropy and coarse-graining of polymer lattice
  models. \emph{Soft Matter} \textbf{2013}, \emph{9}, 11183--11187\relax
\mciteBstWouldAddEndPuncttrue
\mciteSetBstMidEndSepPunct{\mcitedefaultmidpunct}
{\mcitedefaultendpunct}{\mcitedefaultseppunct}\relax
\EndOfBibitem
\bibitem[Ricci and Vergadou(2023)Ricci, and Vergadou]{Ricci2023}
Ricci,~E.; Vergadou,~N. Integrating Machine Learning in the Coarse-Grained
  Molecular Simulation of Polymers. \emph{J. Phys. Chem. B} \textbf{2023},
  \emph{127}, 2302--2322\relax
\mciteBstWouldAddEndPuncttrue
\mciteSetBstMidEndSepPunct{\mcitedefaultmidpunct}
{\mcitedefaultendpunct}{\mcitedefaultseppunct}\relax
\EndOfBibitem
\bibitem[Chennakesavalu \latin{et~al.}(2023)Chennakesavalu, Toomer, and
  Rotskoff]{Rotskoff2023}
Chennakesavalu,~S.; Toomer,~D.~J.; Rotskoff,~G.~M. {Ensuring thermodynamic
  consistency with invertible coarse-graining}. \emph{J. Chem. Phys.}
  \textbf{2023}, \emph{158}, 124126\relax
\mciteBstWouldAddEndPuncttrue
\mciteSetBstMidEndSepPunct{\mcitedefaultmidpunct}
{\mcitedefaultendpunct}{\mcitedefaultseppunct}\relax
\EndOfBibitem
\bibitem[Dunn \latin{et~al.}(2016)Dunn, Foley, and Noid]{Noid2016}
Dunn,~N. J.~H.; Foley,~T.~T.; Noid,~W.~G. Van der Waals Perspective on
  Coarse-Graining: Progress toward Solving Representability and Transferability
  Problems. \emph{Acc. Chem. Res.} \textbf{2016}, \emph{49}, 2832--2840\relax
\mciteBstWouldAddEndPuncttrue
\mciteSetBstMidEndSepPunct{\mcitedefaultmidpunct}
{\mcitedefaultendpunct}{\mcitedefaultseppunct}\relax
\EndOfBibitem
\bibitem[Rosenberger and van~der Vegt(2018)Rosenberger, and van~der
  Vegt]{Rosenberger2018}
Rosenberger,~D.; van~der Vegt,~N. F.~A. Addressing the temperature
  transferability of structure based coarse graining models. \emph{Phys. Chem.
  Chem. Phys.} \textbf{2018}, \emph{20}, 6617--6628\relax
\mciteBstWouldAddEndPuncttrue
\mciteSetBstMidEndSepPunct{\mcitedefaultmidpunct}
{\mcitedefaultendpunct}{\mcitedefaultseppunct}\relax
\EndOfBibitem
\bibitem[Mullinax and Noid(2009)Mullinax, and Noid]{Noid2009}
Mullinax,~J.~W.; Noid,~W.~G. {Extended ensemble approach for deriving
  transferable coarse-grained potentials}. \emph{J. Chem. Phys.} \textbf{2009},
  \emph{131}, 104110\relax
\mciteBstWouldAddEndPuncttrue
\mciteSetBstMidEndSepPunct{\mcitedefaultmidpunct}
{\mcitedefaultendpunct}{\mcitedefaultseppunct}\relax
\EndOfBibitem
\bibitem[Shen \latin{et~al.}(2020)Shen, Sherck, Nguyen, Yoo, Köhler, Speros,
  Delaney, Fredrickson, and Shell]{Shell2020}
Shen,~K.; Sherck,~N.; Nguyen,~M.; Yoo,~B.; Köhler,~S.; Speros,~J.;
  Delaney,~K.~T.; Fredrickson,~G.~H.; Shell,~M.~S. {Learning
  composition-transferable coarse-grained models: Designing external potential
  ensembles to maximize thermodynamic information}. \emph{J. Chem. Phys.}
  \textbf{2020}, \emph{153}, 154116\relax
\mciteBstWouldAddEndPuncttrue
\mciteSetBstMidEndSepPunct{\mcitedefaultmidpunct}
{\mcitedefaultendpunct}{\mcitedefaultseppunct}\relax
\EndOfBibitem
\bibitem[Guenza \latin{et~al.}(2018)Guenza, Dinpajooh, McCarty, and
  Lyubimov]{Guenza2018}
Guenza,~M.~G.; Dinpajooh,~M.; McCarty,~J.; Lyubimov,~I.~Y. Accuracy,
  Transferability, and Efficiency of Coarse-Grained Models of Molecular
  Liquids. \emph{J. Phys. Chem. B} \textbf{2018}, \emph{122},
  10257--10278\relax
\mciteBstWouldAddEndPuncttrue
\mciteSetBstMidEndSepPunct{\mcitedefaultmidpunct}
{\mcitedefaultendpunct}{\mcitedefaultseppunct}\relax
\EndOfBibitem
\bibitem[Craven \latin{et~al.}(2020)Craven, Lubbers, Barros, and
  Tretiak]{craven20c}
Craven,~G.~T.; Lubbers,~N.; Barros,~K.; Tretiak,~S. Machine learning approaches
  for structural and thermodynamic properties of a {Lennard-Jones} fluid.
  \emph{J. Chem. Phys.} \textbf{2020}, \emph{153}, 104502\relax
\mciteBstWouldAddEndPuncttrue
\mciteSetBstMidEndSepPunct{\mcitedefaultmidpunct}
{\mcitedefaultendpunct}{\mcitedefaultseppunct}\relax
\EndOfBibitem
\bibitem[Craven \latin{et~al.}(2020)Craven, Lubbers, Barros, and
  Tretiak]{craven20b}
Craven,~G.~T.; Lubbers,~N.; Barros,~K.; Tretiak,~S. {\it Ex Machina}
  Determination of Structural Correlation Functions. \emph{J. Phys. Chem.
  Lett.} \textbf{2020}, \emph{11}, 4372–4378\relax
\mciteBstWouldAddEndPuncttrue
\mciteSetBstMidEndSepPunct{\mcitedefaultmidpunct}
{\mcitedefaultendpunct}{\mcitedefaultseppunct}\relax
\EndOfBibitem
\bibitem[Rosenberger \latin{et~al.}(2022)Rosenberger, Barros, Germann, and
  Lubbers]{Rosenberger2022}
Rosenberger,~D.; Barros,~K.; Germann,~T.~C.; Lubbers,~N. Machine learning of
  consistent thermodynamic models using automatic differentiation. \emph{Phys.
  Rev. E} \textbf{2022}, \emph{105}, 045301\relax
\mciteBstWouldAddEndPuncttrue
\mciteSetBstMidEndSepPunct{\mcitedefaultmidpunct}
{\mcitedefaultendpunct}{\mcitedefaultseppunct}\relax
\EndOfBibitem
\bibitem[Jin \latin{et~al.}(2020)Jin, Yu, and Voth]{Jin2020transferable}
Jin,~J.; Yu,~A.; Voth,~G.~A. Temperature and Phase Transferable Bottom-up
  Coarse-Grained Models. \emph{J. Chem. Theory Comput.} \textbf{2020},
  \emph{16}, 6823--6842\relax
\mciteBstWouldAddEndPuncttrue
\mciteSetBstMidEndSepPunct{\mcitedefaultmidpunct}
{\mcitedefaultendpunct}{\mcitedefaultseppunct}\relax
\EndOfBibitem
\bibitem[Jin \latin{et~al.}(2019)Jin, Pak, and Voth]{Jin2019entropy}
Jin,~J.; Pak,~A.~J.; Voth,~G.~A. Understanding Missing Entropy in
  Coarse-Grained Systems: Addressing Issues of Representability and
  Transferability. \emph{J. Phys. Chem. Lett.} \textbf{2019}, \emph{10},
  4549--4557\relax
\mciteBstWouldAddEndPuncttrue
\mciteSetBstMidEndSepPunct{\mcitedefaultmidpunct}
{\mcitedefaultendpunct}{\mcitedefaultseppunct}\relax
\EndOfBibitem
\bibitem[Kidder \latin{et~al.}(2021)Kidder, Szukalo, and
  Noid]{Kidder2021energetic}
Kidder,~K.~M.; Szukalo,~R.~J.; Noid,~W. Energetic and entropic considerations
  for coarse-graining. \emph{Eur. Phys. J. B} \textbf{2021}, \emph{94},
  153\relax
\mciteBstWouldAddEndPuncttrue
\mciteSetBstMidEndSepPunct{\mcitedefaultmidpunct}
{\mcitedefaultendpunct}{\mcitedefaultseppunct}\relax
\EndOfBibitem
\bibitem[Pretti and Shell(2021)Pretti, and Shell]{Pretti2021transferable}
Pretti,~E.; Shell,~M.~S. {A microcanonical approach to temperature-transferable
  coarse-grained models using the relative entropy}. \emph{J. Chem. Phys.}
  \textbf{2021}, \emph{155}, 094102\relax
\mciteBstWouldAddEndPuncttrue
\mciteSetBstMidEndSepPunct{\mcitedefaultmidpunct}
{\mcitedefaultendpunct}{\mcitedefaultseppunct}\relax
\EndOfBibitem
\bibitem[Harrison \latin{et~al.}(2018)Harrison, Schall, Maskey, Mikulski,
  Knippenberg, and Morrow]{harrison2018review}
Harrison,~J.~A.; Schall,~J.~D.; Maskey,~S.; Mikulski,~P.~T.;
  Knippenberg,~M.~T.; Morrow,~B.~H. {Review of force fields and intermolecular
  potentials used in atomistic computational materials research}. \emph{Applied
  Physics Reviews} \textbf{2018}, \emph{5}, 031104\relax
\mciteBstWouldAddEndPuncttrue
\mciteSetBstMidEndSepPunct{\mcitedefaultmidpunct}
{\mcitedefaultendpunct}{\mcitedefaultseppunct}\relax
\EndOfBibitem
\bibitem[Kulichenko \latin{et~al.}(2021)Kulichenko, Smith, Nebgen, Li, Fedik,
  Boldyrev, Lubbers, Barros, and Tretiak]{Kulichenko2021}
Kulichenko,~M.; Smith,~J.~S.; Nebgen,~B.; Li,~Y.~W.; Fedik,~N.;
  Boldyrev,~A.~I.; Lubbers,~N.; Barros,~K.; Tretiak,~S. The Rise of Neural
  Networks for Materials and Chemical Dynamics. \emph{The Journal of Physical
  Chemistry Letters} \textbf{2021}, \emph{12}, 6227--6243, PMID: 34196559\relax
\mciteBstWouldAddEndPuncttrue
\mciteSetBstMidEndSepPunct{\mcitedefaultmidpunct}
{\mcitedefaultendpunct}{\mcitedefaultseppunct}\relax
\EndOfBibitem
\bibitem[Smith \latin{et~al.}(2017)Smith, Isayev, and Roitberg]{Smith2017a}
Smith,~J.~S.; Isayev,~O.; Roitberg,~A.~E. ANI-1: an extensible neural network
  potential with DFT accuracy at force field computational cost. \emph{Chem.
  Sci.} \textbf{2017}, \emph{8}, 3192--3203\relax
\mciteBstWouldAddEndPuncttrue
\mciteSetBstMidEndSepPunct{\mcitedefaultmidpunct}
{\mcitedefaultendpunct}{\mcitedefaultseppunct}\relax
\EndOfBibitem
\bibitem[Thaler \latin{et~al.}(2022)Thaler, Stupp, and Zavadlav]{cg-nn-thaler}
Thaler,~S.; Stupp,~M.; Zavadlav,~J. {Deep coarse-grained potentials via
  relative entropy minimization}. \emph{J. Chem. Phys.} \textbf{2022},
  \emph{157}, 244103\relax
\mciteBstWouldAddEndPuncttrue
\mciteSetBstMidEndSepPunct{\mcitedefaultmidpunct}
{\mcitedefaultendpunct}{\mcitedefaultseppunct}\relax
\EndOfBibitem
\bibitem[Wang \latin{et~al.}(2019)Wang, Olsson, Wehmeyer, P{\'e}rez, Charron,
  De~Fabritiis, No{\'e}, and Clementi]{Wang2019machine}
Wang,~J.; Olsson,~S.; Wehmeyer,~C.; P{\'e}rez,~A.; Charron,~N.~E.;
  De~Fabritiis,~G.; No{\'e},~F.; Clementi,~C. Machine learning of
  coarse-grained molecular dynamics force fields. \emph{ACS Central Science}
  \textbf{2019}, \emph{5}, 755--767\relax
\mciteBstWouldAddEndPuncttrue
\mciteSetBstMidEndSepPunct{\mcitedefaultmidpunct}
{\mcitedefaultendpunct}{\mcitedefaultseppunct}\relax
\EndOfBibitem
\bibitem[Fu \latin{et~al.}(2022)Fu, Xie, Rebello, Olsen, and
  Jaakkola]{Fu2022MLCG}
Fu,~X.; Xie,~T.; Rebello,~N.~J.; Olsen,~B.~D.; Jaakkola,~T. Simulate
  time-integrated coarse-grained molecular dynamics with geometric machine
  learning. \emph{arXiv preprint arXiv:2204.10348} \textbf{2022}, \relax
\mciteBstWouldAddEndPunctfalse
\mciteSetBstMidEndSepPunct{\mcitedefaultmidpunct}
{}{\mcitedefaultseppunct}\relax
\EndOfBibitem
\bibitem[Husic \latin{et~al.}(2020)Husic, Charron, Lemm, Wang, Pérez,
  Majewski, Krämer, Chen, Olsson, de~Fabritiis, Noé, and Clementi]{Husic2020}
Husic,~B.~E.; Charron,~N.~E.; Lemm,~D.; Wang,~J.; Pérez,~A.; Majewski,~M.;
  Krämer,~A.; Chen,~Y.; Olsson,~S.; de~Fabritiis,~G.; Noé,~F.; Clementi,~C.
  {Coarse graining molecular dynamics with graph neural networks}. \emph{J.
  Chem. Phys.} \textbf{2020}, \emph{153}, 194101\relax
\mciteBstWouldAddEndPuncttrue
\mciteSetBstMidEndSepPunct{\mcitedefaultmidpunct}
{\mcitedefaultendpunct}{\mcitedefaultseppunct}\relax
\EndOfBibitem
\bibitem[Loose \latin{et~al.}(2023)Loose, Sahrmann, Qu, and
  Voth]{Loose2023_CGNN}
Loose,~T.~D.; Sahrmann,~P.~G.; Qu,~T.~S.; Voth,~G.~A. Coarse-Graining with
  Equivariant Neural Networks: A Path Toward Accurate and Data-Efficient
  Models. \emph{J. Phys. Chem. B} \textbf{2023}, \emph{127}, 10564--10572\relax
\mciteBstWouldAddEndPuncttrue
\mciteSetBstMidEndSepPunct{\mcitedefaultmidpunct}
{\mcitedefaultendpunct}{\mcitedefaultseppunct}\relax
\EndOfBibitem
\bibitem[Zhang \latin{et~al.}(2018)Zhang, Han, Wang, Car, and
  E]{Zhang2018DeePCG}
Zhang,~L.; Han,~J.; Wang,~H.; Car,~R.; E,~W. {DeePCG: Constructing
  coarse-grained models via deep neural networks}. \emph{J. Chem. Phys.}
  \textbf{2018}, \emph{149}, 034101\relax
\mciteBstWouldAddEndPuncttrue
\mciteSetBstMidEndSepPunct{\mcitedefaultmidpunct}
{\mcitedefaultendpunct}{\mcitedefaultseppunct}\relax
\EndOfBibitem
\bibitem[Ruza \latin{et~al.}(2020)Ruza, Wang, Schwalbe-Koda, Axelrod, Harris,
  and Gómez-Bombarelli]{Ruza2020CGionic}
Ruza,~J.; Wang,~W.; Schwalbe-Koda,~D.; Axelrod,~S.; Harris,~W.~H.;
  Gómez-Bombarelli,~R. {Temperature-transferable coarse-graining of ionic
  liquids with dual graph convolutional neural networks}. \emph{J. Chem. Phys.}
  \textbf{2020}, \emph{153}, 164501\relax
\mciteBstWouldAddEndPuncttrue
\mciteSetBstMidEndSepPunct{\mcitedefaultmidpunct}
{\mcitedefaultendpunct}{\mcitedefaultseppunct}\relax
\EndOfBibitem
\bibitem[Duschatko \latin{et~al.}(2024)Duschatko, Vandermause, Molinari, and
  Kozinsky]{Duschatko2024uncertainty}
Duschatko,~B.~R.; Vandermause,~J.; Molinari,~N.; Kozinsky,~B. Uncertainty
  driven active learning of coarse grained free energy models. \emph{npj
  Computational Materials} \textbf{2024}, \emph{10}, 9\relax
\mciteBstWouldAddEndPuncttrue
\mciteSetBstMidEndSepPunct{\mcitedefaultmidpunct}
{\mcitedefaultendpunct}{\mcitedefaultseppunct}\relax
\EndOfBibitem
\bibitem[Airas \latin{et~al.}(2023)Airas, Ding, and Zhang]{Airas2023}
Airas,~J.; Ding,~X.; Zhang,~B. Transferable Implicit Solvation via Contrastive
  Learning of Graph Neural Networks. \emph{ACS Central Science} \textbf{2023},
  \emph{9}, 2286--2297\relax
\mciteBstWouldAddEndPuncttrue
\mciteSetBstMidEndSepPunct{\mcitedefaultmidpunct}
{\mcitedefaultendpunct}{\mcitedefaultseppunct}\relax
\EndOfBibitem
\bibitem[Ge \latin{et~al.}(2023)Ge, Zhang, and Lei]{Ge2023MLcoarse}
Ge,~P.; Zhang,~L.; Lei,~H. {Machine learning assisted coarse-grained molecular
  dynamics modeling of meso-scale interfacial fluids}. \emph{J. Chem. Phys.}
  \textbf{2023}, \emph{158}, 064104\relax
\mciteBstWouldAddEndPuncttrue
\mciteSetBstMidEndSepPunct{\mcitedefaultmidpunct}
{\mcitedefaultendpunct}{\mcitedefaultseppunct}\relax
\EndOfBibitem
\bibitem[Larini \latin{et~al.}(2010)Larini, Lu, and Voth]{Larini2010threebody}
Larini,~L.; Lu,~L.; Voth,~G.~A. {The multiscale coarse-graining method. VI.
  Implementation of three-body coarse-grained potentials}. \emph{J. Chem.
  Phys.} \textbf{2010}, \emph{132}, 164107\relax
\mciteBstWouldAddEndPuncttrue
\mciteSetBstMidEndSepPunct{\mcitedefaultmidpunct}
{\mcitedefaultendpunct}{\mcitedefaultseppunct}\relax
\EndOfBibitem
\bibitem[Das and Andersen(2012)Das, and Andersen]{Das2012threemody}
Das,~A.; Andersen,~H.~C. {The multiscale coarse-graining method. IX. A general
  method for construction of three body coarse-grained force fields}. \emph{J.
  Chem. Phys.} \textbf{2012}, \emph{136}, 194114\relax
\mciteBstWouldAddEndPuncttrue
\mciteSetBstMidEndSepPunct{\mcitedefaultmidpunct}
{\mcitedefaultendpunct}{\mcitedefaultseppunct}\relax
\EndOfBibitem
\bibitem[Wang \latin{et~al.}(2021)Wang, Charron, Husic, Olsson, Noé, and
  Clementi]{Wang2021coarsegrained}
Wang,~J.; Charron,~N.; Husic,~B.; Olsson,~S.; Noé,~F.; Clementi,~C.
  {Multi-body effects in a coarse-grained protein force field}. \emph{J. Chem.
  Phys.} \textbf{2021}, \emph{154}, 164113\relax
\mciteBstWouldAddEndPuncttrue
\mciteSetBstMidEndSepPunct{\mcitedefaultmidpunct}
{\mcitedefaultendpunct}{\mcitedefaultseppunct}\relax
\EndOfBibitem
\bibitem[Noid \latin{et~al.}(2008)Noid, Chu, Ayton, Krishna, Izvekov, Voth,
  Das, and Andersen]{noid2008multiscale1}
Noid,~W.~G.; Chu,~J.-W.; Ayton,~G.~S.; Krishna,~V.; Izvekov,~S.; Voth,~G.~A.;
  Das,~A.; Andersen,~H.~C. The multiscale coarse-graining method. {I}. {A}
  rigorous bridge between atomistic and coarse-grained models. \emph{J. Chem.
  Phys.} \textbf{2008}, \emph{128}\relax
\mciteBstWouldAddEndPuncttrue
\mciteSetBstMidEndSepPunct{\mcitedefaultmidpunct}
{\mcitedefaultendpunct}{\mcitedefaultseppunct}\relax
\EndOfBibitem
\bibitem[Lubbers \latin{et~al.}(2018)Lubbers, Smith, and Barros]{hipnn}
Lubbers,~N.; Smith,~J.~S.; Barros,~K. {Hierarchical modeling of molecular
  energies using a deep neural network}. \emph{J. Chem. Phys.} \textbf{2018},
  \emph{148}, 241715\relax
\mciteBstWouldAddEndPuncttrue
\mciteSetBstMidEndSepPunct{\mcitedefaultmidpunct}
{\mcitedefaultendpunct}{\mcitedefaultseppunct}\relax
\EndOfBibitem
\bibitem[Chigaev \latin{et~al.}(2023)Chigaev, Smith, Anaya, Nebgen,
  Bettencourt, Barros, and Lubbers]{hipnn-ts}
Chigaev,~M.; Smith,~J.~S.; Anaya,~S.; Nebgen,~B.; Bettencourt,~M.; Barros,~K.;
  Lubbers,~N. {Lightweight and effective tensor sensitivity for atomistic
  neural networks}. \emph{J. Chem. Phys.} \textbf{2023}, \emph{158},
  184108\relax
\mciteBstWouldAddEndPuncttrue
\mciteSetBstMidEndSepPunct{\mcitedefaultmidpunct}
{\mcitedefaultendpunct}{\mcitedefaultseppunct}\relax
\EndOfBibitem
\bibitem[Peng \latin{et~al.}(2023)Peng, Pak, Durumeric, Sahrmann, Mani, Jin,
  Loose, Beiter, and Voth]{peng2023openmscg}
Peng,~Y.; Pak,~A.~J.; Durumeric,~A.~E.; Sahrmann,~P.~G.; Mani,~S.; Jin,~J.;
  Loose,~T.~D.; Beiter,~J.; Voth,~G.~A. OpenMSCG: A Software Tool for Bottom-Up
  Coarse-Graining. \emph{J. Phys. Chem. B} \textbf{2023}, \emph{127},
  8537--8550\relax
\mciteBstWouldAddEndPuncttrue
\mciteSetBstMidEndSepPunct{\mcitedefaultmidpunct}
{\mcitedefaultendpunct}{\mcitedefaultseppunct}\relax
\EndOfBibitem
\bibitem[Noid \latin{et~al.}(2007)Noid, Chu, Ayton, and
  Voth]{noid2007multiscale}
Noid,~W.; Chu,~J.-W.; Ayton,~G.~S.; Voth,~G.~A. Multiscale coarse-graining and
  structural correlations: Connections to liquid-state theory. \emph{J. Phys.
  Chem. B} \textbf{2007}, \emph{111}, 4116--4127\relax
\mciteBstWouldAddEndPuncttrue
\mciteSetBstMidEndSepPunct{\mcitedefaultmidpunct}
{\mcitedefaultendpunct}{\mcitedefaultseppunct}\relax
\EndOfBibitem
\bibitem[Rudzinski and Noid(2014)Rudzinski, and
  Noid]{rudzinski2014investigation}
Rudzinski,~J.~F.; Noid,~W.~G. Investigation of Coarse-Grained Mappings via an
  Iterative Generalized Yvon–Born–Green Method. \emph{The Journal of
  Physical Chemistry B} \textbf{2014}, \emph{118}, 8295--8312, PMID:
  24684663\relax
\mciteBstWouldAddEndPuncttrue
\mciteSetBstMidEndSepPunct{\mcitedefaultmidpunct}
{\mcitedefaultendpunct}{\mcitedefaultseppunct}\relax
\EndOfBibitem
\bibitem[Ciccotti \latin{et~al.}(2005)Ciccotti, Kapral, and
  Vanden-Eijnden]{ciccotti2005bluemoon}
Ciccotti,~G.; Kapral,~R.; Vanden-Eijnden,~E. Blue Moon Sampling, Vectorial
  Reaction Coordinates, and Unbiased Constrained Dynamics. \emph{ChemPhysChem}
  \textbf{2005}, \emph{6}, 1809--1814\relax
\mciteBstWouldAddEndPuncttrue
\mciteSetBstMidEndSepPunct{\mcitedefaultmidpunct}
{\mcitedefaultendpunct}{\mcitedefaultseppunct}\relax
\EndOfBibitem
\bibitem[Lubbers(2024)]{hippynn-repo}
Lubbers,~N. e.~a. hippynn. \url{https://github.com/lanl/hippynn}, 2024\relax
\mciteBstWouldAddEndPuncttrue
\mciteSetBstMidEndSepPunct{\mcitedefaultmidpunct}
{\mcitedefaultendpunct}{\mcitedefaultseppunct}\relax
\EndOfBibitem
\bibitem[Smith \latin{et~al.}(2020)Smith, Lubbers, Thompson, and
  Barros]{smith2020simple}
Smith,~J.~S.; Lubbers,~N.; Thompson,~A.~P.; Barros,~K. Simple and efficient
  algorithms for training machine learning potentials to force data.
  \emph{arXiv preprint arXiv:2006.05475} \textbf{2020}, \relax
\mciteBstWouldAddEndPunctfalse
\mciteSetBstMidEndSepPunct{\mcitedefaultmidpunct}
{}{\mcitedefaultseppunct}\relax
\EndOfBibitem
\bibitem[Husic \latin{et~al.}(2020)Husic, Charron, Lemm, Wang, Pérez,
  Majewski, Krämer, Chen, Olsson, de~Fabritiis, Noé, and
  Clementi]{husic2020coarse}
Husic,~B.~E.; Charron,~N.~E.; Lemm,~D.; Wang,~J.; Pérez,~A.; Majewski,~M.;
  Krämer,~A.; Chen,~Y.; Olsson,~S.; de~Fabritiis,~G.; Noé,~F.; Clementi,~C.
  {Coarse graining molecular dynamics with graph neural networks}. \emph{The
  Journal of Chemical Physics} \textbf{2020}, \emph{153}, 194101\relax
\mciteBstWouldAddEndPuncttrue
\mciteSetBstMidEndSepPunct{\mcitedefaultmidpunct}
{\mcitedefaultendpunct}{\mcitedefaultseppunct}\relax
\EndOfBibitem
\bibitem[Fellman \latin{et~al.}(2024)Fellman, Byggmästar, Granberg, Nordlund,
  and Djurabekova]{fellman2024fast}
Fellman,~A.; Byggmästar,~J.; Granberg,~F.; Nordlund,~K.; Djurabekova,~F. Fast
  and accurate machine-learned interatomic potentials for large-scale
  simulations of Cu, Al and Ni. 2024;
  \url{https://arxiv.org/abs/2408.15779}\relax
\mciteBstWouldAddEndPuncttrue
\mciteSetBstMidEndSepPunct{\mcitedefaultmidpunct}
{\mcitedefaultendpunct}{\mcitedefaultseppunct}\relax
\EndOfBibitem
\bibitem[Byggm\"astar \latin{et~al.}(2019)Byggm\"astar, Hamedani, Nordlund, and
  Djurabekova]{byggmastar2019machine}
Byggm\"astar,~J.; Hamedani,~A.; Nordlund,~K.; Djurabekova,~F. Machine-learning
  interatomic potential for radiation damage and defects in tungsten.
  \emph{Phys. Rev. B} \textbf{2019}, \emph{100}, 144105\relax
\mciteBstWouldAddEndPuncttrue
\mciteSetBstMidEndSepPunct{\mcitedefaultmidpunct}
{\mcitedefaultendpunct}{\mcitedefaultseppunct}\relax
\EndOfBibitem
\bibitem[ope()]{open-mscg-tutorial}
MSCG Tutorials: Lesson 01. Force-Matching Single-Site Methanol.
  \url{https://software.rcc.uchicago.edu/mscg/tutorials/lesson-01/README.html},
  Accessed: 2024-03-04\relax
\mciteBstWouldAddEndPuncttrue
\mciteSetBstMidEndSepPunct{\mcitedefaultmidpunct}
{\mcitedefaultendpunct}{\mcitedefaultseppunct}\relax
\EndOfBibitem
\bibitem[Lindahl \latin{et~al.}(2001)Lindahl, Hess, and Van
  Der~Spoel]{lindahl2001gromacs}
Lindahl,~E.; Hess,~B.; Van Der~Spoel,~D. GROMACS 3.0: a package for molecular
  simulation and trajectory analysis. \emph{Molecular modeling annual}
  \textbf{2001}, \emph{7}, 306--317\relax
\mciteBstWouldAddEndPuncttrue
\mciteSetBstMidEndSepPunct{\mcitedefaultmidpunct}
{\mcitedefaultendpunct}{\mcitedefaultseppunct}\relax
\EndOfBibitem
\bibitem[Jorgensen \latin{et~al.}(1996)Jorgensen, Maxwell, and
  Tirado-Rives]{OPLSAA}
Jorgensen,~W.~L.; Maxwell,~D.~S.; Tirado-Rives,~J. Development and Testing of
  the OPLS All-Atom Force Field on Conformational Energetics and Properties of
  Organic Liquids. \emph{J. Am. Chem. Soc.} \textbf{1996}, \emph{118},
  11225--11236\relax
\mciteBstWouldAddEndPuncttrue
\mciteSetBstMidEndSepPunct{\mcitedefaultmidpunct}
{\mcitedefaultendpunct}{\mcitedefaultseppunct}\relax
\EndOfBibitem
\bibitem[Thompson \latin{et~al.}(2022)Thompson, Aktulga, Berger, Bolintineanu,
  Brown, Crozier, in~'t Veld, Kohlmeyer, Moore, Nguyen, Shan, Stevens,
  Tranchida, Trott, and Plimpton]{lammps}
Thompson,~A.~P.; Aktulga,~H.~M.; Berger,~R.; Bolintineanu,~D.~S.; Brown,~W.~M.;
  Crozier,~P.~S.; in~'t Veld,~P.~J.; Kohlmeyer,~A.; Moore,~S.~G.;
  Nguyen,~T.~D.; Shan,~R.; Stevens,~M.~J.; Tranchida,~J.; Trott,~C.;
  Plimpton,~S.~J. {LAMMPS} - a flexible simulation tool for particle-based
  materials modeling at the atomic, meso, and continuum scales. \emph{Comp.
  Phys. Comm.} \textbf{2022}, \emph{271}, 108171\relax
\mciteBstWouldAddEndPuncttrue
\mciteSetBstMidEndSepPunct{\mcitedefaultmidpunct}
{\mcitedefaultendpunct}{\mcitedefaultseppunct}\relax
\EndOfBibitem
\bibitem[Schmid \latin{et~al.}(2011)Schmid, Eichenberger, Choutko, Riniker,
  Winger, Mark, and van Gunsteren]{gromos-54A7}
Schmid,~N.; Eichenberger,~A.~P.; Choutko,~A.; Riniker,~S.; Winger,~M.;
  Mark,~A.~E.; van Gunsteren,~W.~F. Definition and testing of the GROMOS
  force-field versions 54A7 and 54B7. \emph{European Biophysics Journal}
  \textbf{2011}, 843–856\relax
\mciteBstWouldAddEndPuncttrue
\mciteSetBstMidEndSepPunct{\mcitedefaultmidpunct}
{\mcitedefaultendpunct}{\mcitedefaultseppunct}\relax
\EndOfBibitem
\bibitem[Malde \latin{et~al.}(2011)Malde, Zuo, Breeze, Stroet, Poger, Nair,
  Oostenbrink, and Mark]{atb1}
Malde,~A.~K.; Zuo,~L.; Breeze,~M.; Stroet,~M.; Poger,~D.; Nair,~P.~C.;
  Oostenbrink,~C.; Mark,~A.~E. An Automated Force Field Topology Builder (ATB)
  and Repository: Version 1.0. \emph{J. Chem. Theory Comput.} \textbf{2011},
  \emph{7}, 4026--4037\relax
\mciteBstWouldAddEndPuncttrue
\mciteSetBstMidEndSepPunct{\mcitedefaultmidpunct}
{\mcitedefaultendpunct}{\mcitedefaultseppunct}\relax
\EndOfBibitem
\bibitem[Stroet \latin{et~al.}(2018)Stroet, Caron, Visscher, Geerke, Malde, and
  Mark]{atb3}
Stroet,~M.; Caron,~B.; Visscher,~K.~M.; Geerke,~D.~P.; Malde,~A.~K.;
  Mark,~A.~E. Automated Topology Builder Version 3.0: Prediction of Solvation
  Free Enthalpies in Water and Hexane. \emph{J. Chem. Theory Comput.}
  \textbf{2018}, \emph{14}, 5834--5845\relax
\mciteBstWouldAddEndPuncttrue
\mciteSetBstMidEndSepPunct{\mcitedefaultmidpunct}
{\mcitedefaultendpunct}{\mcitedefaultseppunct}\relax
\EndOfBibitem
\bibitem[Martínez \latin{et~al.}(2009)Martínez, Andrade, Birgin, and
  Martínez]{packmol}
Martínez,~L.; Andrade,~R.; Birgin,~E.~G.; Martínez,~J.~M. PACKMOL: A package
  for building initial configurations for molecular dynamics simulations.
  \emph{J. Comput. Chem.} \textbf{2009}, \emph{30}, 2157--2164\relax
\mciteBstWouldAddEndPuncttrue
\mciteSetBstMidEndSepPunct{\mcitedefaultmidpunct}
{\mcitedefaultendpunct}{\mcitedefaultseppunct}\relax
\EndOfBibitem
\bibitem[Jewett \latin{et~al.}(2021)Jewett, Stelter, Lambert, Saladi, Roscioni,
  Ricci, Autin, Maritan, Bashusqeh, Keyes, Dame, Shea, Jensen, and
  Goodsell]{moltemplate}
Jewett,~A.~I.; Stelter,~D.; Lambert,~J.; Saladi,~S.~M.; Roscioni,~O.~M.;
  Ricci,~M.; Autin,~L.; Maritan,~M.; Bashusqeh,~S.~M.; Keyes,~T.; Dame,~R.~T.;
  Shea,~J.-E.; Jensen,~G.~J.; Goodsell,~D.~S. Moltemplate: A Tool for
  Coarse-Grained Modeling of Complex Biological Matter and Soft Condensed
  Matter Physics. \emph{J. Mol. Biol.} \textbf{2021}, \emph{433}, 166841,
  Computation Resources for Molecular Biology\relax
\mciteBstWouldAddEndPuncttrue
\mciteSetBstMidEndSepPunct{\mcitedefaultmidpunct}
{\mcitedefaultendpunct}{\mcitedefaultseppunct}\relax
\EndOfBibitem
\bibitem[Rosenberger \latin{et~al.}(2020)Rosenberger, Lubbers, and
  Germann]{Rosenberger2020}
Rosenberger,~D.; Lubbers,~N.; Germann,~T.~C. {Evaluating diffusion and the
  thermodynamic factor for binary ionic mixtures}. \emph{Physics of Plasmas}
  \textbf{2020}, \emph{27}, 102705\relax
\mciteBstWouldAddEndPuncttrue
\mciteSetBstMidEndSepPunct{\mcitedefaultmidpunct}
{\mcitedefaultendpunct}{\mcitedefaultseppunct}\relax
\EndOfBibitem
\bibitem[Trokhymchuk \latin{et~al.}(2005)Trokhymchuk, Nezbeda, Jirsák, and
  Henderson]{Trokhymchuk2005}
Trokhymchuk,~A.; Nezbeda,~I.; Jirsák,~J.; Henderson,~D. Hard-sphere radial
  distribution function again. \emph{J. Chem. Phys.} \textbf{2005}, \emph{123},
  024501\relax
\mciteBstWouldAddEndPuncttrue
\mciteSetBstMidEndSepPunct{\mcitedefaultmidpunct}
{\mcitedefaultendpunct}{\mcitedefaultseppunct}\relax
\EndOfBibitem
\bibitem[Hansen and McDonald(2006)Hansen, and McDonald]{hansen06}
Hansen,~J.~P.; McDonald,~I.~R. \emph{Theory of simple liquids}; Academic press,
  2006\relax
\mciteBstWouldAddEndPuncttrue
\mciteSetBstMidEndSepPunct{\mcitedefaultmidpunct}
{\mcitedefaultendpunct}{\mcitedefaultseppunct}\relax
\EndOfBibitem
\bibitem[Gray and Gubbins(1984)Gray, and Gubbins]{GrayG1984}
Gray,~C.~G.; Gubbins,~K.~E. \emph{Theory of molecular fluids}; Clarendon Press
  Oxford, 1984; Vol.~1\relax
\mciteBstWouldAddEndPuncttrue
\mciteSetBstMidEndSepPunct{\mcitedefaultmidpunct}
{\mcitedefaultendpunct}{\mcitedefaultseppunct}\relax
\EndOfBibitem
\bibitem[Foidl(1986)]{Foidl1986exact}
Foidl,~C. Exact pair distribution function and structure factor for a
  one-dimensional hard rod mixture. \emph{J. Chem. Phys.} \textbf{1986},
  \emph{85}, 410--417\relax
\mciteBstWouldAddEndPuncttrue
\mciteSetBstMidEndSepPunct{\mcitedefaultmidpunct}
{\mcitedefaultendpunct}{\mcitedefaultseppunct}\relax
\EndOfBibitem
\bibitem[Krämer \latin{et~al.}(2023)Krämer, Durumeric, Charron, Chen,
  Clementi, and Noé]{Kramer2023}
Krämer,~A.; Durumeric,~A. E.~P.; Charron,~N.~E.; Chen,~Y.; Clementi,~C.;
  Noé,~F. Statistically Optimal Force Aggregation for Coarse-Graining
  Molecular Dynamics. \emph{The Journal of Physical Chemistry Letters}
  \textbf{2023}, \emph{14}, 3970--3979, PMID: 37079800\relax
\mciteBstWouldAddEndPuncttrue
\mciteSetBstMidEndSepPunct{\mcitedefaultmidpunct}
{\mcitedefaultendpunct}{\mcitedefaultseppunct}\relax
\EndOfBibitem
\bibitem[Izvekov and Voth(2005)Izvekov, and Voth]{izvekov2005multiscale1}
Izvekov,~S.; Voth,~G.~A. A multiscale coarse-graining method for biomolecular
  systems. \emph{J. Phys. Chem. B} \textbf{2005}, \emph{109}, 2469--2473\relax
\mciteBstWouldAddEndPuncttrue
\mciteSetBstMidEndSepPunct{\mcitedefaultmidpunct}
{\mcitedefaultendpunct}{\mcitedefaultseppunct}\relax
\EndOfBibitem
\bibitem[Noid \latin{et~al.}(2008)Noid, Liu, Wang, Chu, Ayton, Izvekov,
  Andersen, and Voth]{noid2008multiscale2}
Noid,~W.; Liu,~P.; Wang,~Y.; Chu,~J.-W.; Ayton,~G.~S.; Izvekov,~S.;
  Andersen,~H.~C.; Voth,~G.~A. The multiscale coarse-graining method. {II}.
  {N}umerical implementation for coarse-grained molecular models. \emph{J.
  Chem. Phys.} \textbf{2008}, \emph{128}\relax
\mciteBstWouldAddEndPuncttrue
\mciteSetBstMidEndSepPunct{\mcitedefaultmidpunct}
{\mcitedefaultendpunct}{\mcitedefaultseppunct}\relax
\EndOfBibitem
\bibitem[Durumeric \latin{et~al.}(2024)Durumeric, Chen, Noé, and
  Clementi]{durumeric2024learning}
Durumeric,~A. E.~P.; Chen,~Y.; Noé,~F.; Clementi,~C. Learning data efficient
  coarse-grained molecular dynamics from forces and noise. 2024;
  \url{https://arxiv.org/abs/2407.01286}\relax
\mciteBstWouldAddEndPuncttrue
\mciteSetBstMidEndSepPunct{\mcitedefaultmidpunct}
{\mcitedefaultendpunct}{\mcitedefaultseppunct}\relax
\EndOfBibitem
\bibitem[Loose \latin{et~al.}(2023)Loose, Sahrmann, Qu, and
  Voth]{loose2023coarse}
Loose,~T.~D.; Sahrmann,~P.~G.; Qu,~T.~S.; Voth,~G.~A. Coarse-Graining with
  Equivariant Neural Networks: A Path Toward Accurate and Data-Efficient
  Models. \emph{The Journal of Physical Chemistry B} \textbf{2023}, \emph{127},
  10564--10572, PMID: 38033234\relax
\mciteBstWouldAddEndPuncttrue
\mciteSetBstMidEndSepPunct{\mcitedefaultmidpunct}
{\mcitedefaultendpunct}{\mcitedefaultseppunct}\relax
\EndOfBibitem
\bibitem[Jin \latin{et~al.}(2023)Jin, Schweizer, and
  Voth]{jin2023understanding}
Jin,~J.; Schweizer,~K.~S.; Voth,~G.~A. {Understanding dynamics in
  coarse-grained models. I. Universal excess entropy scaling relationship}.
  \emph{The Journal of Chemical Physics} \textbf{2023}, \emph{158},
  034103\relax
\mciteBstWouldAddEndPuncttrue
\mciteSetBstMidEndSepPunct{\mcitedefaultmidpunct}
{\mcitedefaultendpunct}{\mcitedefaultseppunct}\relax
\EndOfBibitem
\bibitem[Izvekov and Voth(2006)Izvekov, and Voth]{izvekov2006modeling}
Izvekov,~S.; Voth,~G.~A. {Modeling real dynamics in the coarse-grained
  representation of condensed phase systems}. \emph{The Journal of Chemical
  Physics} \textbf{2006}, \emph{125}, 151101\relax
\mciteBstWouldAddEndPuncttrue
\mciteSetBstMidEndSepPunct{\mcitedefaultmidpunct}
{\mcitedefaultendpunct}{\mcitedefaultseppunct}\relax
\EndOfBibitem
\bibitem[Hijón \latin{et~al.}(2010)Hijón, Español, Vanden-Eijnden, and
  Delgado-Buscalioni]{hijon2010mori}
Hijón,~C.; Español,~P.; Vanden-Eijnden,~E.; Delgado-Buscalioni,~R.
  Mori–Zwanzig formalism as a practical computational tool. \emph{Faraday
  Discuss.} \textbf{2010}, \emph{144}, 301--322\relax
\mciteBstWouldAddEndPuncttrue
\mciteSetBstMidEndSepPunct{\mcitedefaultmidpunct}
{\mcitedefaultendpunct}{\mcitedefaultseppunct}\relax
\EndOfBibitem
\bibitem[Fritz \latin{et~al.}(2011)Fritz, Koschke, Harmandaris, van~der Vegt,
  and Kremer]{fritz2011multiscale}
Fritz,~D.; Koschke,~K.; Harmandaris,~V.~A.; van~der Vegt,~N. F.~A.; Kremer,~K.
  Multiscale modeling of soft matter: scaling of dynamics. \emph{Phys. Chem.
  Chem. Phys.} \textbf{2011}, \emph{13}, 10412--10420\relax
\mciteBstWouldAddEndPuncttrue
\mciteSetBstMidEndSepPunct{\mcitedefaultmidpunct}
{\mcitedefaultendpunct}{\mcitedefaultseppunct}\relax
\EndOfBibitem
\bibitem[Kinjo and Hyodo(2007)Kinjo, and Hyodo]{kinjo2007equation}
Kinjo,~T.; Hyodo,~S.-a. Equation of motion for coarse-grained simulation based
  on microscopic description. \emph{Phys. Rev. E} \textbf{2007}, \emph{75},
  051109\relax
\mciteBstWouldAddEndPuncttrue
\mciteSetBstMidEndSepPunct{\mcitedefaultmidpunct}
{\mcitedefaultendpunct}{\mcitedefaultseppunct}\relax
\EndOfBibitem
\bibitem[Lei \latin{et~al.}(2010)Lei, Caswell, and Karniadakis]{lei2010direct}
Lei,~H.; Caswell,~B.; Karniadakis,~G.~E. Direct construction of mesoscopic
  models from microscopic simulations. \emph{Phys. Rev. E} \textbf{2010},
  \emph{81}, 026704\relax
\mciteBstWouldAddEndPuncttrue
\mciteSetBstMidEndSepPunct{\mcitedefaultmidpunct}
{\mcitedefaultendpunct}{\mcitedefaultseppunct}\relax
\EndOfBibitem
\bibitem[Gao and Fang(2011)Gao, and Fang]{fao2011semi}
Gao,~L.; Fang,~W. {Semi-bottom-up coarse graining of water based on microscopic
  simulations}. \emph{The Journal of Chemical Physics} \textbf{2011},
  \emph{135}, 184101\relax
\mciteBstWouldAddEndPuncttrue
\mciteSetBstMidEndSepPunct{\mcitedefaultmidpunct}
{\mcitedefaultendpunct}{\mcitedefaultseppunct}\relax
\EndOfBibitem
\bibitem[Davtyan \latin{et~al.}(2015)Davtyan, Dama, Voth, and
  Andersen]{davtyan2015dynamic}
Davtyan,~A.; Dama,~J.~F.; Voth,~G.~A.; Andersen,~H.~C. {Dynamic force matching:
  A method for constructing dynamical coarse-grained models with realistic time
  dependence}. \emph{The Journal of Chemical Physics} \textbf{2015},
  \emph{142}, 154104\relax
\mciteBstWouldAddEndPuncttrue
\mciteSetBstMidEndSepPunct{\mcitedefaultmidpunct}
{\mcitedefaultendpunct}{\mcitedefaultseppunct}\relax
\EndOfBibitem
\bibitem[Davtyan \latin{et~al.}(2016)Davtyan, Voth, and
  Andersen]{davtyan2016dynamic}
Davtyan,~A.; Voth,~G.~A.; Andersen,~H.~C. {Dynamic force matching: Construction
  of dynamic coarse-grained models with realistic short time dynamics and
  accurate long time dynamics}. \emph{The Journal of Chemical Physics}
  \textbf{2016}, \emph{145}, 224107\relax
\mciteBstWouldAddEndPuncttrue
\mciteSetBstMidEndSepPunct{\mcitedefaultmidpunct}
{\mcitedefaultendpunct}{\mcitedefaultseppunct}\relax
\EndOfBibitem
\bibitem[Izvekov(2017)]{izvekov2017microscopic}
Izvekov,~S. {Microscopic derivation of particle-based coarse-grained dynamics:
  Exact expression for memory function}. \emph{The Journal of Chemical Physics}
  \textbf{2017}, \emph{146}, 124109\relax
\mciteBstWouldAddEndPuncttrue
\mciteSetBstMidEndSepPunct{\mcitedefaultmidpunct}
{\mcitedefaultendpunct}{\mcitedefaultseppunct}\relax
\EndOfBibitem
\bibitem[Jung \latin{et~al.}(2017)Jung, Hanke, and Schmid]{jung2017iterative}
Jung,~G.; Hanke,~M.; Schmid,~F. Iterative Reconstruction of Memory Kernels.
  \emph{Journal of Chemical Theory and Computation} \textbf{2017}, \emph{13},
  2481--2488, PMID: 28505440\relax
\mciteBstWouldAddEndPuncttrue
\mciteSetBstMidEndSepPunct{\mcitedefaultmidpunct}
{\mcitedefaultendpunct}{\mcitedefaultseppunct}\relax
\EndOfBibitem
\bibitem[Han \latin{et~al.}(2021)Han, Jin, and Voth]{han2021constructing}
Han,~Y.; Jin,~J.; Voth,~G.~A. {Constructing many-body dissipative particle
  dynamics models of fluids from bottom-up coarse-graining}. \emph{The Journal
  of Chemical Physics} \textbf{2021}, \emph{154}, 084122\relax
\mciteBstWouldAddEndPuncttrue
\mciteSetBstMidEndSepPunct{\mcitedefaultmidpunct}
{\mcitedefaultendpunct}{\mcitedefaultseppunct}\relax
\EndOfBibitem
\bibitem[Izvekov(2017)]{izvekov2017mori}
Izvekov,~S. Mori-Zwanzig theory for dissipative forces in coarse-grained
  dynamics in the Markov limit. \emph{Phys. Rev. E} \textbf{2017}, \emph{95},
  013303\relax
\mciteBstWouldAddEndPuncttrue
\mciteSetBstMidEndSepPunct{\mcitedefaultmidpunct}
{\mcitedefaultendpunct}{\mcitedefaultseppunct}\relax
\EndOfBibitem
\bibitem[Duschatko \latin{et~al.}(2024)Duschatko, Fu, Owen, Xie, Musaelian,
  Jaakkola, and Kozinsky]{duschatko2024}
Duschatko,~B.~R.; Fu,~X.; Owen,~C.; Xie,~Y.; Musaelian,~A.; Jaakkola,~T.;
  Kozinsky,~B. Thermodynamically Informed Multimodal Learning of
  High-Dimensional Free Energy Models in Molecular Coarse Graining. 2024\relax
\mciteBstWouldAddEndPuncttrue
\mciteSetBstMidEndSepPunct{\mcitedefaultmidpunct}
{\mcitedefaultendpunct}{\mcitedefaultseppunct}\relax
\EndOfBibitem
\end{mcitethebibliography}
